\documentclass{article}

\usepackage{amssymb}
\usepackage{mathtools}
\usepackage{amsthm}
\usepackage{latexsym}
\usepackage[pdftex]{graphicx}
\usepackage[dvipsnames]{xcolor}
\usepackage{booktabs}
\usepackage{almostfull}

\usepackage[authoryear,round,longnamesfirst]{natbib}

\usepackage[capitalize]{cleveref}

\Crefname{figure}{Figure}{Figures}

\crefformat{equation}{\textup{#2(#1)#3}}
\crefrangeformat{equation}{\textup{#3(#1)#4--#5(#2)#6}}
\crefmultiformat{equation}{\textup{#2(#1)#3}}{ and \textup{#2(#1)#3}}
{, \textup{#2(#1)#3}}{, and \textup{#2(#1)#3}}
\crefrangemultiformat{equation}{\textup{#3(#1)#4--#5(#2)#6}}%
{ and \textup{#3(#1)#4--#5(#2)#6}}{, \textup{#3(#1)#4--#5(#2)#6}}{, and \textup{#3(#1)#4--#5(#2)#6}}

\Crefformat{equation}{#2Equation~\textup{(#1)}#3}
\Crefrangeformat{equation}{Equations~\textup{#3(#1)#4--#5(#2)#6}}
\Crefmultiformat{equation}{Equations~\textup{#2(#1)#3}}{ and \textup{#2(#1)#3}}
{, \textup{#2(#1)#3}}{, and \textup{#2(#1)#3}}
\Crefrangemultiformat{equation}{Equations~\textup{#3(#1)#4--#5(#2)#6}}%
{ and \textup{#3(#1)#4--#5(#2)#6}}{, \textup{#3(#1)#4--#5(#2)#6}}{, and \textup{#3(#1)#4--#5(#2)#6}}

\crefdefaultlabelformat{#2\textup{#1}#3}

\usepackage{chngcntr}

\newtheorem{theorem}{Theorem}
\usepackage{aliascnt}
\newaliascnt{lemma}{theorem}
\newtheorem{lemma}[lemma]{Lemma}
\aliascntresetthe{lemma}
\newaliascnt{proposition}{theorem}
\newtheorem{proposition}[proposition]{Proposition}
\aliascntresetthe{proposition}
\newaliascnt{conjecture}{theorem}

\aliascntresetthe{conjecture}
\newaliascnt{corollary}{theorem}

\aliascntresetthe{corollary}
\crefname{lemma}{Lemma}{Lemmas}
\crefname{proposition}{Proposition}{Propositions}
\crefname{conjecture}{Conjecture}{Conjectures}
\crefname{corollary}{Corollary}{Corollaries}
\crefname{thm}{Proposition}{Propositions}
\crefname{remark}{Remark}{Remarks}
\crefname{definition}{Definition}{Definitions}

\makeatletter
\newenvironment{subtheorem}[1]{%
  \def\subtheoremcounter{#1}%
  \refstepcounter{#1}%
  \protected@edef\theparentnumber{\csname the#1\endcsname}%
  \setcounter{parentnumber}{\value{#1}}%
  \setcounter{#1}{0}%
  \expandafter\def\csname the#1\endcsname{\theparentnumber.\alph{#1}}%
  \ignorespaces
}{%
  \setcounter{\subtheoremcounter}{\value{parentnumber}}%
  \ignorespacesafterend
}
\makeatother
\newcounter{parentnumber}
\newtheorem{thm}{Proposition}
\theoremstyle{definition}
\newtheorem{definition}[theorem]{Definition}
\newtheorem{remark}[theorem]{Remark}

\usepackage{mpcsymbols}

\newcommand{\td}{\tilde{d}}

\newcommand{\ulambda}{\underline{\lambda}}

\newcommand{\newinf}{\mathop{\mathrm{inf}\vphantom{\mathrm{sup}}}}

\title{Background results for robust minmax control \\of linear dynamical
 systems\thanks{An earlier version of these results was presented at the short course
``Robust Nonlinear Model Predictive Control: Recent Advances in Design and Computation,'' University of California, Santa Barbara, CA, March 25--28, 2024.}}
\author{
 James B. Rawlings, %\thanks{} {} and
 Davide Mannini, and  %\thanks{} {} and
 Steven J. Kuntz %\thanks{} \\
\thanks{The authors gratefully acknowledge the financial support of the
		National Science Foundation (NSF) under Grant Nos. 2027091 and 2138985.
\texttt{jbraw@ucsb.edu}, \texttt{dmannini@ucsb.edu}, \texttt{skuntz@ucsb.edu}.}\\%
 Department of Chemical Engineering\\
 University of California, Santa Barbara}

\date{\today}

\begin{document}

\maketitle

The purpose of this note is to summarize the arguments required to derive the results appearing in robust minmax control of linear dynamical systems using a quadratic stage cost. The main result required in robust minmax control is
% \cref{cor:conminmax} and \cref{prop:conminmaxlin}.
\cref{prop:conminmaxalt}.
Moreover, the solution to the trust-region problem given in \cref{prop:conquad} and \cref{lem:dLquad} may be of more general interest.

This note has been revised twice.  The second version corrected the optimal $u$ and $w$ formulas and combined several results; this third version corrects two typos in the statement of \cref{prop:conminmaxalt} and adds \cref{lem:Ldual} and \cref{prop:ineqdual}.  Theorem numbering is stable across all three versions.  The appendix lists the changes in detail.

The authors acknowledge Claude (Anthropic), which reviewed the second version line by line, checked the matrix identities numerically, and drafted the corrections listed in the appendix.

\section*{Linear algebra}

We assume throughout that the parameters $D \in \bbR^{n\times n}$ is symmetric, $A \in \bbR^{m\times n}, b \in \bbR^{m}, d \in \bbR^n$ or $d \in \bbR^{n+m}$.  Let $A^+ \in \bbR^{n\times m}$ denote the pseudoinverse of matrix $A \in \bbR^{m \times n}$. Let $N(A)$ and $R(A)$ denoted the null space and range space of matrix $A$, respectively. 
Let $\norm{A}$ denote the spectral norm of matrix $A$, i.e., its largest singular value.  Every matrix whose norm appears in what follows is symmetric and positive semidefinite, for which the spectral norm coincides with the largest eigenvalue.
We will also make use of the singular value decomposition (SVD) of  $A$ given by  
\begin{equation}
A=\begin{bmatrix}U_1 & U_2 \end{bmatrix} 
\begin{bmatrix} \Sigma_r & 0 \\ 0 & 0 \end{bmatrix}
\begin{bmatrix}V_1' \\ V_2' \end{bmatrix} = U_1 \Sigma_r V_1'
\label{eq:SVDA}
\end{equation}
and $r$ is the rank of $A$.
The properties of the SVD and the fundamental theorem of linear algebra imply that the orthonormal columns of $U_1$ and $U_2$ are bases for $R(A)$ and $N(A')$, respectively, and the orthonormal columns of $V_1$ and $V_2$ are bases for $R(A')$ and $N(A)$, respectively.\footnote{Edge cases:  $A=0$ has $r=0$ and empty $U_1, V_1, \Sigma_r$ matrices, so $U=U_2, V=V_2$ and $R(A)=\{0\}, R(A')=\{0\}, N(A) = \bbR^n, N(A')=\bbR^m$. At the other extreme, if $A$ is square and invertible, $r=m=n$ and $U_2, V_2$ are empty so $U=U_1, V=V_1$, and $R(A)=\bbR^n, R(A')=\bbR^m, N(A)=\{0\}, N(A') = \{0\}$.}  We also have that $A^+ = V_1 \Sigma_r^{-1}U_1'$.

First, we require solutions to linear algebra problems when such solutions exist.
\begin{proposition}[Solving linear algebra problems.]
\label{prop:linalg}
Consider the linear algebra problem
\begin{equation*}
Ax=b
\end{equation*}
\begin{enumerate}
\item A solution exists if and only if $b \in R(A)$.
\item For $b \in R(A)$, the solution (set of solutions) is given by\footnote{We overload the addition symbol 
to mean set addition when adding singletons ($A^+b$) and sets ($N(A)$).}
\begin{equation}
x^0 \in A^+ b + N(A)
\label{eq:linalg}
\end{equation}
\end{enumerate}
\end{proposition}
\begin{proof}
By definition of range, if $b \notin R(A)$ there is no $x$ such that $Ax=b$, and if $b \in R(A)$, there is an $x$ such that $Ax=b$, which is the same as the existence condition.  
For $b \in R(A)$, let $z \in \bbR^n$ denote a value so that $Az=b$, and let $q$ be an arbitrary element in $N(A)$ so $x^0 = A^+b + q$. To show \cref{eq:linalg} are solutions, note that 
\begin{equation*}
Ax^0 = A (A^+b + q) = AA^+Az = Az = b
\end{equation*}
where we have used the definition of the null space and one of the pseudoinverse's defining properties,  $AA^+A=A$.  To show that \cref{eq:linalg} are all solutions, let $x'$ denote a solution. We then have $A(x' - A^+b) = b-b=0$, so $x' - A^+b \in N(A)$ or $x' \in A^+b + N(A)$. Since $x'$ is an arbitrary solution, \cref{eq:linalg} gives all solutions. 
\end{proof}
Note that if one is interested in \textit{deriving} \cref{eq:linalg} rather than establishing that it is correct as we did here, use the two orthogonal coordinate systems provided by the SVD of $A$, and let $x = V \alpha$, $b= U \beta$, and solve that simpler \textit{decoupled} linear algebra problem for $\alpha^0$ as a function of $\beta$, and convert back to $x^0$ in terms of $b$. 

If $b \notin R(A)$, $x^0$ is still well-defined, but $Ax^0-b = (AA^+ - I)b = -U_2U_2'b \neq 0$.
In this case, the $x^0$ given in \cref{eq:linalg} solves $\min_x \norm{Ax-b}$ (least-squares solution), and achieves value $\norm{Ax^0-b} = \norm{U_2'b}$.  

\paragraph{Positive semidefinite matrices.}
We say that a matrix $M \in \bbR^{n\times n}$ is positive semidefinite, denoted $M \geq 0$, if $M$ is symmetric and $x'Mx \geq 0 $ for all $x  \in \bbR^n$.

\section*{Optimization}

We shall appeal without proof to one theorem for existence of solutions to optimization problems, the Weierstrass (extreme value) theorem. It says that a continuous function on a closed and bounded set attains its min and max on the set.\footnote{Proofs for the multivariate version required here can be found in \citet[p. 198]{mangasarian:1994}, \citet[Corollary 5.1.25]{polak:1997}, \citet[p. 11]{rockafellar:wets:1998}, and \citet[Proposition A.7]{rawlings:mayne:diehl:2020}.}  As we specialize to the results of interest in this note, next we consider convex, differentiable functions.
\begin{definition}[Convex function]
\label{def:convex}
A function $V : \bbR^n \rightarrow \bbR$ is convex if
\begin{equation}
 V(\alpha u + (1-\alpha) v) \leq \alpha V(u) + (1-\alpha) V(v)
\label{eq:convex}
\end{equation} 
\end{definition} 
for all $u, v \in \bbR^n$ and $0 \leq \alpha  \leq 1$.

If the function $V$ is differentiable, then it is convex if and only if
\begin{equation}
 V(v)  \geq V(u) + (v-u)' \frac{d V}{du}(u)
\label{eq:lowerbound}
\end{equation} 
for all $u, v \in \bbR^n$.  See \citet[pp.69--70]{boyd:vandenberghe:2004} for a proof of this fact.

An immediate consequence of this global lower bound is that $u^0$ is a minimizer  of $V$ if and only if $(dV/du)(u^0) = 0$.
\begin{proposition}
\label{prop:zeroderiv}
A convex, differentiable function $V : \bbR^n \rightarrow \bbR$ has a minimizer $u^0$ if and only if 
$(dV/du) (u^0) = 0$.
\end{proposition} 
\begin{proof}
To establish sufficiency, assume $(dV/du)(u^0) = 0$; \cref{eq:lowerbound} then implies $V(v) \geq V(u^0)$ for all $v \in \bbR^n$, and therefore $u^0$ is the minimizer of $V$.

To establish necessity, assume that $u^0$ is optimal but that, contrary to what is to be proven, $(dV/du)(u^0) \neq 0$, and let $h = -(dV/du)(u^0)$ so that  the directional derivative satisfies
\begin{equation*}
 \lim_{\lambda \rightarrow 0^+} \frac{V(u^0 + \lambda h) - V(u^0)}{\lambda} = h' \frac{dV}{du}(u^0) = -\norm{\frac{dV}{du} (u^0)}^2
\end{equation*} 
Given this limit, for every $\epsilon>0$ there exists $\delta(\epsilon)>0$ such that
\begin{equation*}
\frac{V(u^0 + \lambda h) - V(u^0)}{\lambda} \leq -\norm{\frac{dV}{du} (u^0)}^2 + \epsilon
\end{equation*} 
for all $0 < \lambda \leq \delta$.  Choose  $\epsilon = (1/2) \norm{(dV/du)(u^0)}^2 > 0$, and we have that
\begin{equation*}
V(u^0 + \lambda h) \leq V(u^0) -(\lambda/2)\norm{\frac{dV}{du} (u^0)}^2
\end{equation*} 
for $0 < \lambda \leq \delta$.  This inequality contradicts the optimality of $u^0$ and, therefore $(dV/du)(u^0)=0$, which establishes necessity, and the proposition is proven.
\end{proof}

When considering robust control of linear dynamical systems with quadratic stage cost, quadratic functions play a central role. We have the following result about their convexity.
\begin{proposition}[Convex quadratic functions]
\label{prop:convquad}
The quadratic function $V(u) = (1/2)u'Du + u'd + c$ is convex if and only if $D \geq 0$.
\end{proposition} 
\begin{proof}
We establish that the quadratic term $f(u) \eqbyd u'Du$ is convex by substituting $\alpha u + (1-\alpha)v$ into function $f$ and rearranging the terms
\begin{equation*}
f(\alpha u + (1-\alpha) v) - \big( \alpha f(u) + (1-\alpha) f(v) \big) = -\alpha(1-\alpha)(u-v)'D(u-v)
\end{equation*} 
Since $-\alpha(1-\alpha)<0$ for $\alpha \in (0,1)$, we have that the right-hand side is less than or equal to zero for every $u,v \in \bbR^n$ if and only if $D \geq 0$, verifying \eqref{eq:convex} for the function $f$. 
 
It is then straightforward to show that the linear function $u'd$ and the constant function $c$ are both convex
by directly verifying \eqref{eq:convex}. It is also straightforward to establish that linear combinations of convex functions are convex by verifying \eqref{eq:convex}, and, therefore the function $V$ is convex if and only if $D \geq 0$.
\end{proof} 

The following optimization result for convex quadratic functions is then useful in the ensuing discussion.
\begin{proposition}[Minimum of quadratic functions]
\label{prop:min}
Consider the quadratic function $V(\cdot): \bbR^{n} \rightarrow \bbR$
with $D \in \bbR^{n\times n}$
\begin{equation*}
V(u) \eqbyd \half u'Du + u'd
\end{equation*}
\begin{enumerate}
\item 
A solution to $\min_u V$ exists if and only if  $D \geq 0$ and $d \in R(D)$.
\item For $d \in R(D)$, the minimizer and optimal value function are
\begin{equation}
u^0 \in -D^+d + N(D) \qquad V^0 = -(1/2) d'D^+ d
\label{eq:minsoln}
\end{equation}
and $(d / d u) V(u)= 0$ at $u^0$.
\end{enumerate}
\end{proposition} 
\begin{proof}
Forward implication.  Assume $D \geq 0$ and $d \in R(D)$. 
The function $V(u)$ is differentiable and convex (\cref{prop:convquad}) so from \cref{prop:zeroderiv}, a solution exists if and only if the derivative is zero.  Taking the derivative gives $(d/du) V(u) = Du + d$.  From \cref{prop:linalg}, $Du+d = 0$ has a solution and the set of all solutions is $u^0 \in -D^+d + N(D)$; evaluating $V$ at the solution and noting that $d'N(D)=0$ since $d \in R(D)$ gives $V(u^0) = -(1/2)d'D^+d$ establishing \cref{eq:minsoln}.

Reverse implication. Assume $D$ is not positive semidefinite. Since $D$ is symmetric, it has an eigenvalue decomposition $D=Q \Lambda Q'$ where $Q$ is orthonormal and $\Lambda$ is diagonal and real, and has at least one negative eigenvalue, $\lambda_i$ with corresponding eigenvector $q_i$. Letting $u= \alpha q_i$ gives cost $V(u) = (1/2) \alpha^2 \lambda_i + \alpha q_i'd$. A solution does not exist because $\lim_{\alpha \rightarrow \infty} V(u) = - \infty$. Next assume that $D\geq 0$ but $d \notin R(D)$. Then there is no solution to $Du + d = 0$, and a solution does not exist by \cref{prop:zeroderiv}. The reverse implication has been established and the proof is complete.
\end{proof} 
For maximization problems, we can replace $D \geq 0$ with $D \leq 0$ and min with max.

Next we add a linear equality constraint to \cref{prop:min} and develop the following result.
\begin{subtheorem}{thm}
\setcounter{parentnumber}{\value{theorem}-1}
\end{subtheorem} 
\begin{subtheorem}{thm}
\begin{thm}[Minimum of semidefinite quadratic functions subject to linear constraints]
\label{prop:mina}
Consider the quadratic function $V(\cdot): \bbR^{n} \rightarrow \bbR$
with $D \in \bbR^{n\times n} \geq 0$
\begin{equation*}
V(u) \eqbyd \half u'Du + u'd
\end{equation*}
Let $A \in \bbR^{m\times n}, b \in \bbR^m$ and consider the linear constraint $Au = b$.
\begin{enumerate}
\item 
A solution to $\min_u V(u)$ s.t. $Au=b$ exists if and only if $c \in R(M)$ where
\begin{gather*}
c \eqbyd \begin{bmatrix} d \\ b \end{bmatrix} \qquad
M \eqbyd \begin{bmatrix} D & -A' \\ -A & 0 \end{bmatrix}
\end{gather*} 
\item If a solution exists, the minimizer and optimal value function are given by
\begin{align}
\begin{bmatrix} u \\ \gamma \end{bmatrix}^0 &\in -M^+c + N(M) \label{eq:minsola}\\
V^0 &= -(1/2) c'M^+ c \label{eq:mincosta}
\end{align} 
where $\gamma \in \bbR^m$ is a convenient auxiliary variable.
\end{enumerate}
\end{thm}
% \end{subtheorem} 
\begin{proof}
An optimal $u$ exists if and only if $Au=b$ and $V(u) \leq V(u+p)$ for every $p$ such that $u + p$ is feasible, i.e., $A(u+p) = b$, or $Ap = 0$, or $p \in N(A)$.  Calculating the cost difference gives the condition
\begin{equation*}
V(u+p) - V(u) = p'(Du+d) + (1/2) p'Dp \geq 0 \qquad \text{for all } p \in N(A)
\end{equation*}  
With $D \geq 0$, since $p'Dp$ is quadratic in $p$, and $p'(Du+d)$ is linear in $p$, and $N(A)$ is a linear subspace, this inequality can be satisfied if and only if $p' (Du+d) =  0$, or $Du+d$ is orthogonal to $N(A)$. To see this, assume there exists $p \in N(A)$ such that $p'(Du+d) \neq 0$. Then we can choose a $p \in N(A)$ with the correct sign, to make $p'(Du+d)$ negative, and $p$ small enough so that this linear term dominates the nonnegative quadratic term, and violates the inequality, giving a contradiction. Finally, since $R(A') = N(A)^\perp$, the inequality is satisfied if and only if $(Du+d) \in R(A')$.  We express this condition as $Du + d = A' \gamma$ for $\gamma \in \bbR^m$, and we see the reason to introduce this auxiliary variable.  Combining this  condition with the feasiblity condition gives a solution to 
\begin{equation}
\begin{bmatrix} D & - A' \\ -A & 0 \end{bmatrix}
\begin{bmatrix} u \\ \gamma \end{bmatrix} =
- \begin{bmatrix} d \\ b \end{bmatrix} \qquad 
  M \begin{bmatrix} u \\ \gamma \end{bmatrix} = -c
\label{eq:dLdugam}
\end{equation}   
as the necessary and sufficient condition for existence of an optimal solution.  Applying \cref{prop:linalg} then gives $c\in R(M)$ as the existence condition and \cref{eq:minsola} as the optimizer.
To calculate the optimal cost,  define $L(u,\gamma)$
\begin{equation*}
L(u, \gamma) \eqbyd V(u) - \gamma'(Au - b)
= (1/2) \begin{bmatrix} u \\ \gamma \end{bmatrix}' M \begin{bmatrix} u \\ \gamma \end{bmatrix}
 + \begin{bmatrix} u \\ \gamma \end{bmatrix}' c
\end{equation*}  
This is exactly the same functional form as in \cref{prop:min}, and applying that result for the cost after setting $dL/d(u,\gamma) = 0$, as we have done in \cref{eq:dLdugam}, gives
 \begin{equation*}
L(u^0, \gamma^0) = -(1/2) c'M^+c
\end{equation*} 
Since $Au^0 = b$, we also have that $L(u^0,\gamma^0) = V(u^0)$, which establishes \cref{eq:mincosta} and completes the proof.
\end{proof} 
Although we did not invoke any of the theory of Lagrange multipliers in this development, clearly we see their emergence as a convenient tool for solving the constrained optimization problem. We can further notice that introducing the Lagrangian function $L(u, \gamma)$ enables us to express the existence condition as simply solving
\begin{equation*}
dL(u,\gamma)/ d(u, \gamma) =  M\begin{bmatrix} u\\ \gamma \end{bmatrix} + c = 0
\end{equation*} 
We shall shortly make these connections after introducing minimax problems.

Before we do that, we consider more general quadratic functions.  We know that $D$ is positive semidefinite or positive definite in the robust minmax control problems of interest. But it may prove useful to treat more general cases. Consider the removing the restriction that $D\geq 0$, so that it may be indefinite, i.e., $D$ may have both positive and negative eigenvalues.  This case is not typically of much interest in unconstrained optimization since neither minimization nor maximization problems have solutions.  But \textit{constrained} optimization can have solutions with indefinite $D$. We have the following result.
\begin{thm}[Minimum of general quadratic functions subject to linear constraints]
\label{prop:minb}
Consider the quadratic function $V(\cdot): \bbR^{n} \rightarrow \bbR$
with symmetric $D \in \bbR^{n\times n}$
\begin{equation*}
V(u) \eqbyd \half u'Du + u'd
\end{equation*}
Let $A \in \bbR^{m\times n}, b \in \bbR^m$ and consider the linear constraint $Au = b$.
\begin{enumerate}
\item 
A solution to $\min_u V(u)$ s.t. $Au=b$ exists if and only if
\begin{enumerate}
 \item  $V_2'DV_2 \geq 0$
 \item  $\begin{bmatrix} d \\ b \end{bmatrix} \in R(M) \qquad 
M \eqbyd \begin{bmatrix} D & -A' \\ -A & 0 \end{bmatrix}$
\end{enumerate} 
where the columns of $V_2$ are a basis for $N(A)$, the nullspace of $A$.

\item If a solution exists, the minimizer and optimal value function are given by
\begin{equation*}
\begin{bmatrix} u \\ \lambda \end{bmatrix}^0 = -M^+ \begin{bmatrix} d \\ b \end{bmatrix} + N(M) \qquad
V^0 = -(1/2) \begin{bmatrix} d \\ b \end{bmatrix}' M^+ \begin{bmatrix} d \\ b \end{bmatrix}\\
\end{equation*} 
\end{enumerate}
\end{thm}
\begin{proof}
Consider the SVD of the constraint matrix $A=USV = U_1\Sigma V_1'$, with $A^+=V_1\Sigma^{-1} U_1'$ and the columns of  $V_2$ serving as an orthonormal basis for $N(A)$.  Express the decision variable $u$ in the $V$ coordinate system by, $u=V\alpha = V_1\alpha_1 + V_2\alpha$. To have a solution, the constraint must be feasible, so $b \in R(A)$, which is  given in the bottom half of assumption 1 (b). The set of all feasible $u$ is then given by $u_f = A^+b + V_2\alpha_2$ with $\alpha_2$ an arbitrary vector.
%\begin{enumerate}
%\item
Substituting $u_f$ into the objective function gives
\begin{equation*}
 V(u_f) = (1/2) \alpha_2'(V_2'DV_2)\alpha_2 + \alpha_2' V_2'(DA^+ b + d) + (1/2) b'A'^+DA^+b + d'A^+b
\end{equation*} 
From \cref{prop:min}, the unconstrained $ \min_{\alpha_2}V(u_f)$ has a solution if and only if
(i) $V_2'DV_2 \geq 0$ and (ii) $V_2'(DA^+b +d) \in R(V_2'DV_2)$.  The first is given by assumption 1. (a).
From assumption 1. (b) we have that there exists a $(u,w)$ such that $d = Du -A'w$ and $b = -Au$.  Therefore $V_2'(DA^+b+d) = V_2'(D(I-A^+A) u - A'w) = V_2'DV_2(V_2'u)$, which implies (ii).
Since both (i) and (ii) are satisfied, the optimal $\alpha_2$ is obtained by setting $dV(u_f)/d\alpha_2$ to zero  
giving 
 \begin{equation*}
0 = V_2'DV_2 \alpha_2^0 + V_2'(DA^+b +d)  = V_2'D( V_2\alpha_2^0 + A^+b) + V_2'd 
= V_2'D u^0 + V_2' d
\end{equation*} 
So we have $V_2'D u^0 = - V_2'd$ and $Au^0 = b$ as the two conditions for optimality, which we write as
\begin{equation}
 \begin{bmatrix} V_2' D \\ -A \end{bmatrix} u^0 = -\begin{bmatrix} V_2' d \\ b \end{bmatrix}
 \label{eq:u0}
\end{equation}
We next show that \cref{eq:u0} are equivalent to the stationary conditions
\begin{equation}
 \begin{bmatrix} D & -A' \\ -A & 0 \end{bmatrix} \begin{bmatrix} u\\ \lambda \end{bmatrix}^*
 = - \begin{bmatrix} d \\ b  \end{bmatrix} 
\label{eq:ulam*}
  \end{equation}  
We wish to show that the two solution sets are equal, $u^* = u^0$.  First we show that $u^* \subseteq u^0$. Consider a solution of \cref{eq:ulam*}, take the top equation and multiply by $V_2'$, and note that $V_2'A'=0$ to obtain $V_2'D u^* -V_2'A' \lambda^* = V_2'D u^* = -V_2'd$, which is the top half of \cref{eq:u0}. The bottom half of \cref{eq:ulam*} and \cref{eq:u0} are identical, so we have shown that $u^*$ satisfies \eqref{eq:u0} and therefore $u^* \subseteq u^0$. 
  
Next we show that $u^0 \subseteq u^*$. Consider a solution of \cref{eq:u0} and define
the corresponding $\lambda^0 \eqbyd A'^+(Du^0 + d) +N(A')$.  We then have
$Du^0 - A'\lambda^0 = (D-A'A'^+D)u^0 - A'A'^+d = V_2 V_2' D u^0 - V_1V_1' d = -V_2V_2'd-V_1V_1'd = -d$, and we have satisfied the top half of \eqref{eq:ulam*}.  Again, since the bottom half of \cref{eq:u0} and \cref{eq:ulam*} are identical we have shown that $(u^0, \lambda^0)$ satisfy \eqref{eq:ulam*} and therefore $u^0 \subseteq u^*$. Taken together, we have shown that $u^0 = u^*$.

Given this solution, we then calculate the optimal cost. Noting that $Au^0 = b$ and $u^0=u^*$ gives
\begin{align*}
V(u^0) &= (1/2)u'^0Du^0 +u'^0d - \lambda'^* (Au^0-b) =
(1/2) \begin{bmatrix} u \\ \lambda \end{bmatrix}'^* M  \begin{bmatrix} u \\ \lambda \end{bmatrix}^* +
 \begin{bmatrix} u \\ \lambda \end{bmatrix}'^* \begin{bmatrix} d \\ b \end{bmatrix} \\
 &= (1/2)  \begin{bmatrix} u \\ \lambda \end{bmatrix}'^* \begin{bmatrix} d \\ b \end{bmatrix}
 = -(1/2)  \begin{bmatrix} d \\ b \end{bmatrix}' M^+  \begin{bmatrix} d \\ b \end{bmatrix}
\end{align*} 
where in the last equality we also use the fact that $(d,b) \perp N(M)$ since $(d,b) \in R(M)$. 
   
  \end{proof}

 If instead of using the Lagrange multiplier approach, we eliminate the constraint $Ax=b$ and solve the remaining optimization problem, we obtain the following very different looking result.
\begin{thm}[Minimum of general quadratic functions subject to linear constraints (Alternate)]
\label{prop:minc}
Consider the quadratic function $V(\cdot): \bbR^{n} \rightarrow \bbR$
with symmetric $D \in \bbR^{n\times n}$
\begin{equation*}
V(u) \eqbyd \half u'Du + u'd
\end{equation*}
Let $A \in \bbR^{m\times n}, b \in \bbR^m$ and consider the linear constraint $Au = b$.
\begin{enumerate}
\item 
A solution to $\min_u V(u)$ s.t. $Au=b$ exists if and only if
\begin{enumerate}
 \item  $V_2'DV_2 \geq 0$
\item  $\begin{bmatrix} V_2' & V_2'DA^+ \\ 0 & I \end{bmatrix} \begin{bmatrix} d \\ b \end{bmatrix}
\in R \bigg( \begin{bmatrix}V_2'DV_2 & 0 \\ 0 & A \end{bmatrix} \bigg)$
% \item  $\begin{bmatrix} V_2' & V_2'DA^+ \end{bmatrix} \begin{bmatrix} d \\ b \end{bmatrix}
% \in R(V_2'DV_2) \qquad b \in R(A)$
\end{enumerate} 
where the columns of $V_2$ are a basis for $N(A)$, the nullspace of $A$.

\item If a solution exists, the minimizer and optimal value function are given by
\begin{align}
u^0 &= -\begin{bmatrix} W & (WD-I)A^+ \end{bmatrix} \begin{bmatrix} d \\ b \end{bmatrix}
 + V_2N(V_2'DV_2) \label{eq:minub}\\
V^0 &= -(1/2) \begin{bmatrix} d \\ b \end{bmatrix}' X \begin{bmatrix} d \\ b \end{bmatrix}\label{eq:mincostb}\\
\intertext{with}
W &\eqbyd V_2(V_2'DV_2)^+V_2' \notag\\
 X &\eqbyd 
\begin{bmatrix} W & (WD-I)A^+ \\
A'^+(DW-I) & A'^+(DWD - D)A^+ \notag \end{bmatrix}
\end{align} 
\end{enumerate}
\end{thm}
\begin{proof}\mbox{ }
\begin{enumerate}
\item As in the proof of \cref{prop:minb}, define $u_f = A^+b + V_2\alpha_2$ and substituting $u_f$ into the objective  function gives 
\begin{equation*}
 V(u_f) = (1/2) \alpha_2'(V_2'DV_2)\alpha_2 + \alpha_2' V_2'(DA^+ b + d) + (1/2) b'A'^+DA^+b + d'A^+b
\end{equation*} 
From \cref{prop:min}, $ \min_{\alpha_2}V(u_f)$ has a solution if and only if
 \begin{equation*}
 V_2'DV_2 \geq 0 \qquad V_2'(DA^+b + d) \in R(V_2'DV_2)
\end{equation*} 
% D indefinite here; stop using this argument
% Note that for $D \geq 0$, $R(V_2'DV_2) = N^\perp(V_2'DV_2) = N^\perp(DV_2) = R(V_2'D)$,  
which, along with $b \in R(A)$, establishes the existence condition.

 \item If the existence condition is satisfied, again from \cref{prop:min},  the optimal $\alpha_2$ is given by $\alpha_2^0 = -(V_2'DV_2)^+ V_2'(DA^+b + d) + N(V_2'DV_2)$. Converting back to the $u$ coordinates
%and noting that for $D \geq 0$, $N(V_2'DV_2)= N(V_2'DV_2)$
 gives
\begin{align*}
u^0 &= A^+b + V_2 \alpha_2^0 \\
&= -W d + (A^+ -WDA^+)b + V_2N(V_2'DV_2) \\
u^0 &= -\begin{bmatrix}  W & (WD-I)A^+ \end{bmatrix} \begin{bmatrix} d \\ b \end{bmatrix} + V_2N(V_2'DV_2)
\end{align*} 
which establishes \cref{eq:minub}.  
  
Substituting $\alpha_2^0$ into $V(u_f)$ and using \cref{eq:minsoln} to evaluate the $\alpha_2$ terms gives
\begin{multline*}
V^0 = -(1/2)[V_2'(DA^+b +d)]'(V_2'DV_2)^+[V_2'(DA^+b +d)] + \\
 (1/2) b'A'^+DA^+b + d'A^+b
\end{multline*} 
Putting $(d,b)$ into a vector and factoring gives \cref{eq:mincostb}, which completes the proof.
\end{enumerate} 
\end{proof}

\end{subtheorem} 
 
\paragraph{Partitioned semidefinite matrices.}
We make extensive use of partitioned matrices
\begin{equation*}
M= \begin{bmatrix} M_{11} & M_{12} \\ M'_{12} & M_{22} \end{bmatrix}
\end{equation*}
We have the following result for positive semidefinite partitioned matrices \cite[p.651]{boyd:vandenberghe:2004}.
\begin{proposition}[Positive semidefinite partitioned matrices]
\label{prop:psd}
The matrix $M \geq 0$ if and only if $M_{11} \geq 0$, $M_{22} - M_{12}'M_{11}^+ M_{12} \geq 0$, and $R(M_{12}) \subseteq R(M_{11})$.
\end{proposition}
\begin{proof} \mbox{ }
\begin{enumerate}
\item Forward implication. 
Define $V(x,y) \eqbyd (1/2) (x,y)'M(x,y)$, and assume $M\geq 0$.
%The problem $\min_{x,y} V(x,y)$ has a solution from \cref{prop:min}.
Expanding $V$ using the partitioned matrix
\begin{align}
V(x,y) &= (1/2) \begin{bmatrix} x\\ y \end{bmatrix}' 
\begin{bmatrix} M_{11} & M_{12} \\ M'_{12} & M_{22} \end{bmatrix}
 \begin{bmatrix} x\\ y \end{bmatrix} \notag \\
&= (1/2) \big( x' M_{11} x + 2 x'M_{12}y + y'M_{22} y \big) \label{eq:psdexpand}\\
&\geq 0,  \quad \text{for all } (x,y) \notag
\end{align} 
Setting $y=0$ in \cref{eq:psdexpand} implies that $M_{11} \geq 0$. Since $M_{11}\geq 0$, $V(x,y)$ is a differentiable, convex function of $x$ for any $y$. Therefore $\min_x V(x,y)$ has a solution for every $y$, and \cref{prop:min} then implies $M_{12}y \in R(M_{11})$ for every $y$, which is equivalent to $R(M_{12}) \subseteq R(M_{11})$. Substituting the minimizer over $x$, $x^0 = -M_{11}^+M_{12}y$ into $V$ gives
\begin{equation}
 V(x^0,y)  = (1/2) y' (M_{22} - M'_{12}M_{11}^+ M_{12})  y
\label{eq:schurpsd}
\end{equation} 
and since $V(x,y)  \geq 0$ for all $(x,y)$, we have that $M_{22}-M'_{12}M^+_{11}M_{12}\geq 0$, and the forward implication is established.  

\item Reverse implication. Assume $M_{11}\geq 0$, $M_{22}-M'_{12}M^+_{11}M_{12}\geq 0$, and 
$R(M_{12}) \subseteq R(M_{11})$, and we establish that $M\geq 0$.  For proof by contradiction, assume there exists an $(\ox, \oy)$ such that $(\ox,\oy)'M(\ox,\oy) < 0$.
By \cref{prop:min}, we know that  $\min_{x} V(x,\oy)$ exists since $M_{11}\geq 0$ and $M_{12}\oy \in R(M_{11})$, and it has value $V^0 = V(x^0, \oy)$ with $x^0 = -M_{11}^+ M_{12} \oy$. Substituting this into $V$ gives $V^0 = (1/2) \oy'(M_{22} -M'_{12}M^+_{11}M_{12})\oy \geq 0$ because matrix $M_{22}-M'_{12}M^+_{11}M_{12}\geq 0$. By optimality of $x^0$, $V(x,\oy) \geq V^0 \geq 0$ for all $x$.  But that contradicts $V(\ox, \oy) < 0$, and we  conclude $M \geq 0$, and the proof is complete.  \qedhere
\end{enumerate}
\end{proof}    

Note that 
\begin{equation*}
\begin{bmatrix} M_{11} & M_{12} \\ M'_{12} & M_{22} \end{bmatrix} \geq 0 
\quad \text{if and only if} \quad
\begin{bmatrix} M_{22} & M'_{12} \\ M_{12} & M_{11} \end{bmatrix} \geq 0 
\end{equation*}
So we can also conclude that $M\geq 0$ if and only if
$M_{11} \geq 0$, $M_{22}\geq 0$, $M_{22} - M'_{12}M_{11}^+ M_{12} \geq 0$, 
$M_{11} - M_{12}M_{22}^+ M'_{12} \geq 0$, 
$R(M_{12}) \subseteq R(M_{11})$, and $R(M_{12}') \subseteq R(M_{22})$. Note also that given the partitioning in $M$, we define
\begin{align}
\tM_{11} &\eqbyd M_{22} - M_{12}'M_{11}^+ M_{12} \notag \\
\tM_{22} &\eqbyd M_{11} - M_{12}M_{22}^+ M'_{12}
\label{eq:schurdef}
\end{align} 
and $\tM_{11}$ is known as the Schur complement of $M_{11}$,  and $\tM_{22}$ is known as the Schur complement of $M_{22}$.

\paragraph{Constraints and Lagrangians.}
Next we require a standard optimization result for using a Lagrangian to reformulate a constrained minimization as an unconstrained minmax problem. The following result will be useful for this purpose.
Let $U \subseteq \bbR^n$ be a nonempty compact set and $V(\cdot): U \rightarrow \bbR$  be a continuous function on $U$.
Define the Lagrangian function $L(\cdot): U \times \bbR \rightarrow \bbR$ as
\begin{equation}
L(u, \lambda) = V(u) - \lambda \rho(u, U)
\label{eq:Ldefgen}
\end{equation}
where $\rho(\cdot): \bbR^n \times U \rightarrow \bbR_{\geq 0}$ is any convenient continuous indicator function that evaluates to zero if and only if $u \in U$.  
Denote a minmax problem as
\begin{equation*}
\infp_u \sup_\lambda L(u, \lambda)  
\end{equation*}
When a solution to this problem exists, we define the optimal value $L^*$ and solution set $u^*$ as
\begin{equation*}
L^* = \min_u \max_\lambda L(u, \lambda) \qquad u^* = \arg \min_u \max_\lambda L(u, \lambda)
\end{equation*}
It is convenient in the subsequent development to define the maximizer of the inner problem 
\begin{equation*}
\olambda(u) \eqbyd \arg \max_{\lambda} L(u, \lambda), \quad u \in U
\end{equation*}

\begin{proposition}[Constrained minimization and Lagrangian minmax]
\label{prop:conlag}
Let $U \subseteq \bbR^n$ be a nonempty compact set and $V(\cdot): U \rightarrow \bbR$  be a continuous function on $U$, and $L(\cdot): U \times \bbR \rightarrow \bbR$ be defined as $L(u, \lambda) \eqbyd V(u) - \lambda \rho(u)$, where $\rho(\cdot): \bbR^n \rightarrow \bbR$ is continuous and satisfies $\rho(u) = 0$ if and only if $u \in U$, as in \cref{eq:Ldefgen}.  Note that $\rho(\cdot)$ is \textit{not} required to be nonnegative; the distance from $u$ to $U$ is one admissible choice, and $\rho(u) = \half(u'u-1)$ for $U = \{u \mid u'u=1\}$ is another.
Consider the constrained optimization problem
\begin{equation}
\inf_{u \in U} V(u)
\label{eq:conmin}
\end{equation}
and the (unconstrained) Lagrangian minmax problem
\begin{equation}
\infp_{u} \sup_\lambda L(u,\lambda) 
\label{eq:Lgen}
\end{equation}
\begin{enumerate}
\item Solutions to both problems exist.
\item Let $V^0$ be the solution and $u^0$ be the set of optimizers of $\min_{u\in U} V(u)$. Let $L^*$ be the solution and $u^*$ the set of optimizers of $\min_u \max_\lambda L(u, \lambda)$.  Then 
\begin{equation*}
V^0 = L^* \qquad u^0=u^* \qquad \olambda(u^*) = \bbR
\end{equation*}

\item The same statements hold for the maximization problem $\sup_{u \in U} V(u)$ and its Lagrangian reformulation $\sup_u \inf_\lambda L(u,\lambda)$, for which
\begin{equation*}
\inf_\lambda L(u,\lambda) = \begin{cases} V(u), & \quad u \in U \\ -\infty, & \quad u \notin U \end{cases}
\end{equation*}
\end{enumerate}
\end{proposition}

\begin{proof}
The solution to \cref{eq:conmin} exists by the Weierstrass theorem. Denote the optimal value $V^0$ and solution set $u^0 \subseteq U$ which satisfy $V(u^0)=V^0$. We show that a solution to \cref{eq:Lgen} also exists.  Consider the inner supremum.  From the definitions of functions $L$ and $\rho$, and because $\lambda$ ranges over all of $\bbR$ so that $\sup_\lambda -\lambda \rho(u) = +\infty$ whenever $\rho(u) \neq 0$, regardless of the sign of $\rho(u)$, we conclude 
\begin{equation*}
\sup_\lambda L(u,\lambda) = \begin{cases} 
V(u),  & \quad u \in U \\
+\infty,  & \quad u \notin U
\end{cases}
\end{equation*}
Then consider the outer infimum. We have that 
\begin{equation*}
L^* = \infp_u \sup_\lambda L(u,\lambda) = \inf_{u \in U} V(u) = \min_{u\in U}V(u) = V^0
\end{equation*}
So the solution to \eqref{eq:Lgen} exists with value $L^* = V^0$.  Taking the argument gives
\begin{equation*}
u^* \eqbyd \arg \infp_u \sup_\lambda L(u,\lambda) = \arg \min_{u\in U} V(u) = u^0
\end{equation*}
We note that 
\begin{equation*}
\olambda(u) = \arg \sup_\lambda L(u, \lambda) = \bbR \qquad 
\end{equation*}
For all $u \subseteq U$. Since $u^* = u^0$ and $u^0 \subseteq U$
\begin{equation*}
\olambda(u^*) = \bbR
\end{equation*} 
Statement 3 follows by applying statements 1 and 2 to $-V(\cdot)$ and replacing $\lambda$ by $-\lambda$.
\end{proof}

\section*{Minmax and Maxmin}

More generally, we are interested in a function $V(u,w)$ $V:U \times W \rightarrow \bbR$
and the optimization problems
\begin{gather*}
\newinf_{u \in U} \sup_{w \in W} V(u,w) \qquad 
\sup_{w \in W} \newinf_{u \in U} V(u,w)
\end{gather*}
We assume in the following that the $\inf$ and $\sup$ are achieved on the
respective sets and replace them with $\min$ and $\max$. 

\paragraph{Continuous functions.}
Let's start with the following generalization of von Neumann's minimax theorem.
\begin{theorem}[Minimax Theorem]
\label{th:minimax}
Let $U \subset \bbR^m$ and $W \subset \bbR^n$ be compact convex sets. If 
$V: U \times W \to \bbR$ is a continuous function that is convex-concave, i.e.,
$V(\cdot ,w):U \to \bbR$ is convex for all $w \in W$, and
$V(u, \cdot ):W \to \bbR$ is concave for all $u \in U$\\
Then we have that
\begin{equation*}
\min_{u \in U} \max_{w \in W} V(u,w) = \max_{w \in W} \min_{u \in U} V (u,w) 
\end{equation*}
\end{theorem}

\citet{vonneumann:1928} treats a finite, two-person, zero-sum game:
the decision sets are probability simplices and the payoff is \textit{bilinear}
on them, which is the special case of a convex-concave $V$ in which both
convexity and concavity are affine.
% \citet{ville:1938} and others enlarged the
% sets to subsets of infinite-dimensional linear spaces, but the payoff remained
% linear.
The first minimax theorem for a genuinely convex-concave $V$ appears to
be \citet{shiffman:1949}; the first in the journal literature is
\citet{kneser:1952}.
% , who separates two disjoint convex sets by a hyperplane
% instead of invoking a fixed-point theorem.
% \citet{fan:1953} then weakened
% convexity to convexlike and continuity to semicontinuity in one variable, and
% \citet{nikaido:1954} obtained a quasi-convex-concave version from Brouwer's
% theorem.
\cref{th:minimax} is the specialization to continuous, convex-concave
$V$ on compact convex $U \subset \bbR^m$, $W \subset \bbR^n$ of
\citet[Thm.~3.4]{sion:1958}, which
%unifies these two lines of argument and
is the standard reference; \citet[\S\S33--37]{rockafellar:1970} also provides a self-contained
treatment of convex-concave saddle-point theory in $\bbR^n$.

Note that existence of min and max is guaranteed by compactness of $U,W$ (closed, bounded).  Also note that the following holds for any continuous function $V$
\begin{equation*}
\min_{u \in U} \max_{w \in W} V(u,w) \geq \max_{w \in W} \min_{u \in U} V (u,w) 
\end{equation*}
This is often called \textit{weak duality}. It's easy to establish.  We are regarding the switching of the order of min and max as a form of duality. (Think of observability and controllability as duals of each other.)

So when this inequality achieves equality, that's often called \textit{strong duality.}   So the minimax theorem says that continuous functions that are convex-concave on compact sets satisfy strong duality.  When strong duality is not achieved, we refer to the difference as the \textit{duality gap}, which is positive due to weak duality
\begin{equation*}
\min_{u \in U} \max_{w \in W} V(u,w) - \max_{w \in W} \min_{u \in U} V (u,w)  > 0
\end{equation*}

\paragraph{Saddle Points.}
In characterizing solutions of these problems, it is useful to define a saddle point for the function $V(u,w)$.  
\begin{definition}[Saddle point]
The point (set) $(u^*, w^*) \subseteq U \times W$  is called a saddle point (set) for $V(\cdot)$ if
\begin{equation}
V(u^*,w) \leq V(u^*,w^*) \leq V(u, w^*) \quad \text{for all } u \in U, w \in W
\label{eq:saddle}
\end{equation}
\end{definition}
\begin{proposition}[Saddle-point theorem]
\label{prop:saddle}
The point (set) $(u^*, w^*) \subseteq U\times W$ is a saddle point (set) for function $V(\cdot)$ if and only if strong duality holds and $(u^*,w^*)$ is a solution to the two problems
\begin{gather}
\min_{u \in U} \max_{w \in W} V(u,w) = \max_{w \in W} \min_{u \in U} V(u,w)
= V(u^*,w^*) \\
u^* = \arg \min_{u \in U} \max_{w \in W} V(u,w) \qquad
w^* = \arg \max_{w \in W} \min_{u \in U} V(u,w)
\label{eq:minmaxdual}
\end{gather}
\end{proposition}
In the following development it is convenient to define the
solutions to the minimization and maximization problems
\begin{gather}
\oV(u) \eqbyd \max_{w \in W} V(u, w), 
\quad u \in U \qquad
\uV(w) \eqbyd \min_{u \in U} V(u,w), 
\quad w \in W
\label{eq:inneropt}
\end{gather}
Note that \cref{eq:minmaxdual} implies that $\max_{w\in W}\uV(w) = \uV(w^*)$ and
$\min_{u\in U}\oV(u)= \oV(u^*)$. 

\begin{remark} Note that \cref{eq:minmaxdual} also implies that
\begin{equation}
\max_{w \in W} \min_{u \in U} V(u, w) = \min_{u \in U} V(u,w^*) \qquad 
\min_{u\in U} \max_{w \in W} V(u, w) = \max_{w \in W} V(u^*,w)
\label{eq:minmaxalso}
\end{equation}

To establish this remark, note that
\begin{equation*}
\max_{w \in W} \min_{u \in U} V(u, w) = \max_{w \in W} \uV(w) = 
\uV(w^*) = \min_{u \in U}V(u,w^*)
\end{equation*}
Similarly,
\begin{equation*}
\min_{u\in U} \max_{w \in W} V(u, w) = \min_{u \in U}\oV(u) = 
\oV(u^*) = \max_{w \in W} V(u^*,w)
\end{equation*}
\end{remark}

Next we prove \cref{prop:saddle}
\begin{proof}
First we establish that \cref{eq:minmaxdual} implies \cref{eq:saddle}.
Note that by optimality, the first equality in \cref{eq:minmaxalso}, which is a
consequence of assuming \cref{eq:minmaxdual}, implies that  $V(u^*,w^*) \leq
V(u,w^*)$ for all $u \in U$, and the second implies 
that $V(u^*,w^*) \geq V(u^*,w)$ for all $w \in W$. Taken together, these
are \cref{eq:saddle}. 

Next we show that \cref{eq:saddle} implies \cref{eq:minmaxdual}.
We know that the following holds by weak duality
\begin{equation}
\max_{w \in W} \min_{u \in U} V(u,w) \leq
\min_{u \in U} \max_{w \in W} V(u,w)
\label{eq:weakduality}
\end{equation}
So we wish to show that the reverse inequality also holds to establish strong
duality, i.e., the first equality in \cref{eq:minmaxdual}. To
that end note that from \cref{eq:saddle}
\begin{equation*}
V(u^*,w) \leq V(u,w^*) \quad \text{for all } w \in W, u \in U
\end{equation*}
Since this holds for all $w \in W$, it also holds for a maximizer, and therefore
\begin{equation*}
\max_{w \in W} V(u^*,w) \leq V(u,w^*)  \quad \text{for all } u \in U
\end{equation*}
The left-hand side will not be larger if instead of evaluating at
$u=u^*\in U$, we minimize over all $u \in U$, giving
\begin{equation*}
\min_{u \in U} \max_{w \in W} V(u,w) \leq V(u,w^*)  \quad \text{for all } u \in U
\end{equation*}
Now if this inequality holds for all $u \in U$, it also holds for the
minimizer on the right-hand side  so that
\begin{equation*}
\min_{u \in U} \max_{w \in W} V(u,w) \leq \min_{u \in U} V(u,w^*)
\end{equation*}
We can only increase the value of the right-hand side if instead of
evaluating at $w=w^* \in W$, we maximize over all $w \in W$, giving
\begin{equation*}
\min_{u \in U} \max_{w \in W} V(u,w) \leq \max_{w \in W} \min_{u \in U} V(u,w)
\end{equation*}
Note that this is the weak duality inequality \cref{eq:weakduality} written in
the reverse direction, so combining with weak duality, we have that 
\begin{equation*}
\min_{u \in U} \max_{w \in W} V(u,w) = \max_{w \in W} \min_{u \in U} V(u,w)
\end{equation*}
and strong duality is established.  

We next show that $u^*$ solves the minmax problem.
From the defined optimizations in \cref{eq:inneropt} we have that
\begin{align}
\min_{u \in U} \max_{w \in W} V(u,w) &= \min_{u \in U}
                                             \oV(u) \label{eq:minmax} \\
\max_{w \in W} \min_{u \in U}  V(u,w)  &= \max_{w \in W} \uV(w) \label{eq:maxmin}
\end{align}
Next choose an arbitrary $u_1\in U$ and assume for contradiction that $\oV(u_1) < \oV(u^*)$.  From the definition of $\oV$ we then have that
\begin{equation*}
\max_{w \in W} V(u_1,w) < \max_{w \in W} V(u^*,w)
\end{equation*}
Therefore since $w^* \in W$
\begin{equation*}
V(u_1,w^*) < \max_{w \in W} V(u^*,w)
\end{equation*}
But from the saddle-point condition, \cref{eq:saddle}, $\max_{w \in W} V(u^*,w) \leq V(u,w^*)$ for all $u \in U$, which contradicts the previous inequality since $u_1 \in U$. Therefore $\oV(u_1) \geq \oV(u^*)$, and since $u_1$ is an arbitrary element of $U$, $u^*$ solves the minmax problem \cref{eq:minmax}.

Similarly we can show that $w^*$ solves the maxmin problem \cref{eq:maxmin} by exchanging the variables $w$ and $u$ and the operations $\max$ and $\min$. Therefore $(w^*,u^*)$ solves \cref{eq:minmaxdual}, and we have established that \cref{eq:saddle} implies \cref{eq:minmaxdual}.
\end{proof}

Although $U$ and $W$ are taken to be compact above, the proof of \cref{prop:saddle} uses only weak duality and the existence of the indicated optima; compactness enters only as a convenient sufficient condition for that existence.  We may therefore apply \cref{prop:saddle} to unbounded $U$ and $W$ whenever the optima in question are known to exist, as we do in \cref{lem:Lquad} with $U = \bbR^n$ and $W = \bbR$.

In the following development it is convenient to define the solutions to the inner minimization and maximization problems
\begin{gather}
%\uV(w) \eqbyd \min_{u \in U} V(u,w), \quad  
\uu^0(w) := \arg \min_{u \in U} V(u,w),  \quad w \in W \\
%\oV(u) \eqbyd \max_{w \in W} V(u,w),  \quad 
\ow^0(u) := \arg \max_{w \in W} V(u,w),  \quad u \in U 
\label{eq:inneropt_sol}
\end{gather}
Note that these inner solution sets are too ``large''  in the following sense. Even if we evaluate them at the optimizers of their respective  outer problems, we know only that 
\begin{gather*}
u^* \subseteq \uu^0(w^*)  \qquad  w^* \subseteq \ow^0(u^*)
\end{gather*}
and these subsets may be strict. So we have to exercise some care when we exploit strong duality and want to extract the optimizer from a  dual problem. We shall illustrate this issue in the upcoming results.

\paragraph{Quadratic functions.}

In control problems, we min and max over possibly unbounded sets, so we need something other than compactness to guarantee existence of solutions.  When we have linear dynamic models and quadratic stage cost (LQ), we can use the following results for quadratic functions.

\begin{subtheorem}{thm}
\setcounter{theorem}{12}
\setcounter{parentnumber}{\value{theorem}-1}
\end{subtheorem} 
\begin{subtheorem}{thm}
\begin{thm}[Alternative to 12]
%\begin{proposition}
\label{prop:SPalt}
Consider the quadratic function $V(\cdot): \bbR^{m+n} \rightarrow \bbR$
\begin{equation*}
V(u,w) \eqbyd (1/2) \begin{bmatrix} u \\ w \end{bmatrix}'
\begin{bmatrix} M_{11} & M_{12} \\ M'_{12} & M_{22} \end{bmatrix}
\begin{bmatrix} u \\ w \end{bmatrix} +
\begin{bmatrix} u \\ w \end{bmatrix}' 
\begin{bmatrix} d_1 \\d_2 \end{bmatrix}
\end{equation*}
with $M$ symmetric, $M_{11} \in \bbR^{m\times m}$, $M_{22} \in \bbR^{n\times n}$, $M_{12} \in
\bbR^{m\times n}, d \in \bbR^{m+n}$.

For $d \in R(M)$, define stationary points $(u^*, w^*)$ of function $V(\cdot)$
as $d V(u,w) / d (u,w) = 0$ at $(u^*, w^*)$, satisfying
\begin{equation}
M \begin{bmatrix} u^* \\ w^* \end{bmatrix}  = -d
\qquad 
\begin{bmatrix} u^* \\ w^* \end{bmatrix} \in  -M^+ d + N(M) 
\label{eq:u*w*}
\end{equation} 
 with cost
\begin{equation}
V(u^*, w^*) = - (1/2) d'M^+ d
\label{eq:V*}
\end{equation}
Denote the solutions to the inner optimizations, when they exist, by $\uu^0(w) \eqbyd \arg \min_u V(u,w)$ and $\ow^0(u) \eqbyd \arg \max_w V(u,w)$. 
\begin{enumerate}
%\item  Solutions to $\min_u \max_w V$ exist if and only if  $d \in R(M)$, $M_{22} \leq 0$,
%and $\tM_{22}\geq 0$, (not sure this is true, problem is constrained)
\item  Solutions to $\min_u \max_w V$ exist if $d \in R(M)$, $M_{22} \leq 0$, and $\tM_{22}\geq 0$
and satisfy
% \begin{gather*}
% \arg \min_u \max_w V(u,w) = u^* \qquad \ow^0(u^*) = -M_{22}^+ (M_{12}' u^* + d_2) + N(M_{22}) \\
% V(u^*,\ow^0(u^*)) = - (1/2) d'M^+ d
% \end{gather*} 
\begin{gather*}
\arg \min_u \max_w V(u,w) = u^* \qquad M_{12}'u^* +  M_{22} \ow^0(u^*) = - d_2 \\
V(u^*,\ow^0(u^*)) = - (1/2) d'M^+ d
\end{gather*} 

%% \item  Similarly, solutions to $\max_w \min_u V$ exist if and only if  $d \in R(M)$,  $M_{11} \ge q 0$,
%and $\tM_{11}\leq 0$, (ditto, not sure this is true, problem is constrained)
\item  Similarly, solutions to $\max_w \min_u V$ exist if $d \in R(M)$,  $M_{11} \geq 0$,
and $\tM_{11}\leq 0$, and satisfy
\begin{gather*}
\arg \max_w \min_u V(u,w) = w^* \qquad M_{11} \uu^0(w^*) + M_{12} w^* = -d_1 \\
V(\uu^0(w^*),w^*) = - (1/2) d'M^+ d
\end{gather*} 

\item Strong duality holds if and only if $d \in R(M)$, $M_{11} \geq 0$, and $M_{22} \leq 0$, which gives
\begin{equation*}
\min_u \max_w V(u,w) = \max_w \min_u V(u,w) = V(u^*, w^*)
\end{equation*}
In this case both inner optimizations exist and $u^* \subseteq \uu^0(w^*)$ and $w^* \subseteq \ow^0(u^*)$.  
\end{enumerate}
\end{thm} 
\end{subtheorem} 
\begin{proof}\mbox{ }
\begin{enumerate}
\item 
%By \cref{prop:min} a solution to the inner $\max_wV$ exists if and only if $M_{22}\geq 0$, and $M_{12}'u + d_2= 0$.
First assume $d \in R(M)$ and expand $V(\cdot)$ as
\begin{equation}
V(u,w) = (1/2) w'M_{22}w + w'(M_{12}'u + d_2) + (1/2) u'M_{11}u + u'd_1
\label{eq:Vexpand}
\end{equation} 
From \cref{prop:min}, $\max_wV$ exists if and only $M_{22} \leq 0$ and $M_{12}'u + d_2 \in R(M_{22})$.  This condition is satisfied for some nonempty set of $u$ by the bottom half of $d \in R(M)$. For such $u$ we have the necessary and sufficient condition for the optimum
\begin{equation}
 M_{22}\ow^0 +  M_{12}'u + d_2 = 0
\label{eq:bottom}
\end{equation}
which defines an implicit function $\ow^0(u)$, and optimal value given by \eqref{eq:minsoln}
\begin{align*}
\ow^0(u) &= -M^+_{22} (M_{12}'u + d_2) + N(M_{22}) \\
V(u, \ow^0(u)) &= (1/2) u'\tM_{22}u + u'(d_1 - M_{12}M_{22}^+ d_2) - (1/2) d_2'M_{22}^+d_2
\end{align*} 
% where $\tM_{22}$ is the Schur complement of $M_{22}$ defined in
% \eqref{eq:schurdef}.
% Note that $\tM_{22}\geq 0$ since $M_{11} \geq 0$ and
% $M_{22}\leq 0$, which implies $M_{22}^+ \leq 0$.
To find the optimal $u$, we cannot simply set $(d/du) V(u, \ow^0(u))$
to zero because we require $M_{12}'u + d_2 \in R(M_{22})$ for the
existence of $V(u, \ow^0(u))=\max_wV(u,w)$. To handle this range constraint, we
use a linear equality constraint $M_{12}'u+d_2 = -M_{22}y$ where $y$ is a slack
variable. Under the equality constraint, the problem $\min_u\max_w V$ is
equivalent to the following constrained minimization:
\[
  \min_{u,y} V(u, \ow^0(u)) \qquad \textnormal{subject to} \qquad M_{12}'u+d_2 =
 - M_{22}y.
\]
To solve this we apply \cref{prop:mina}  using
 \begin{equation*}
 D = \begin{bmatrix} \tM_{22} & 0 \\ 0 & 0 \end{bmatrix} \qquad
 d = \begin{bmatrix} d_1 - M_{12}M_{22}^+ d_2 \\ 0 \end{bmatrix} \qquad
 A = \begin{bmatrix} M_{12}' & M_{22} \end{bmatrix} \qquad b = -d_2
\end{equation*} 
From \cref{prop:mina}, a solution to this constrained problem exists if and only if $\tM_{22} \geq 0$ \textit{and} the right-hand side of the optimality conditions below lies in the range of their matrix. The first condition holds by assumption; the second is established at the end of this part.  The necessary and sufficient conditions for the optimal $(u,y,\gamma)$ are
\begin{equation*}
 \begin{bmatrix} \tM_{22} & 0 & -M_{12} \\ 0 & 0 & -M_{22} \\ -M_{12}' & -M_{22} & 0 \end{bmatrix}
 \begin{bmatrix} u \\ y \\ \gamma \end{bmatrix}
 = - \begin{bmatrix} d_1 - M_{12}M_{22}^+ d_2 \\ 0 \\ -d_2 \end{bmatrix}
\end{equation*}  
Replacing the last row with its negative, changing the sign of $\gamma$, and adding $M_{12}M_{22}^+$ times the last row to the first row then gives the equivalent system
\begin{equation*}
\begin{bmatrix} M_{11} & M_{12}M_{22}^+ M_{22} & M_{12} \\ 0 & 0 & M_{22} \\ M_{12}' & M_{22} & 0 \end{bmatrix}
 \begin{bmatrix} u \\ y \\ \gamma \end{bmatrix}
 = - \begin{bmatrix} d_1 \\ 0 \\ d_2 \end{bmatrix}
\end{equation*}  
Finally, we introduce a new unknown $w = M_{22}^+M_{22} y + \gamma$, and switch the second and third equations to obtain
\begin{equation*}
\begin{bmatrix} M_{11} & M_{12} & 0 \\  M_{12}' & M_{22} & 0 \\ 0 & 0 & M_{22} \end{bmatrix}
 \begin{bmatrix} u \\ w \\ \gamma \end{bmatrix}
 = - \begin{bmatrix} d_1 \\ d_2 \\ 0 \end{bmatrix}
\end{equation*}  
The equations for the optimal $(u,w)$ and $\gamma$ are now decoupled. Every step from the optimality conditions to this system is reversible: the row operations are invertible, and given any $(u,w,\gamma)$ solving the decoupled system we recover a solution of the optimality conditions by taking $y = w$ and $\gamma = (I-M_{22}^+M_{22})w$, which satisfies $w = M_{22}^+M_{22}y + \gamma$ and $M_{22}\gamma = 0$.  Solving $\min_u\max_w V$ is therefore \textit{equivalent} to solving the stationary conditions defining the $(u^*,w^*)$ pair, with the auxiliary variable satisfying $\gamma \in N(M_{22})$.  In particular, the range condition deferred above is satisfied precisely when the decoupled system has a solution, which by \cref{prop:linalg} is precisely when $d \in R(M)$, as assumed at the outset; this equivalence also supplies the necessity of $d \in R(M)$ used in part 3.  Also, $w^* \subseteq \ow^0(u^*)$, since from \cref{eq:bottom}, $\ow^0$ is all solutions to just the bottom half of the stationary conditions defining $w^*$. Evaluating the cost for $(u^*,w^*)$ satisfying the stationary conditions gives $V(u^*,w^*)=-(1/2)d'M^+d$, and we have shown that this is the optimal value  for $\min_u \max_w V(u,w)$. 
\item 
To solve the $\max_w \min_u V$ problem, change the sign of $V(u,w)$ and relabel the variables to obtain
\begin{equation*}
\min_u\max_w V(u,w) = - \max_u\min_w -V(u,w) = -\max_w\min_u -V(w,u)
\end{equation*} 
Applying the results of 1. to the last expression produces the results of 2, e.g., $M_{22}\leq 0$ in 1. transforms to $-M_{11} \leq 0$ or $M_{11} \geq 0$ in 2., and so on.
% the previous problem, which pro
%  take the negative of the objective to
% obtain $\max_w\min_u V=-\min_w\max_u(-V)$. Therefore the exact same procedure
% can be used here, and it produces the same solutions and optimal values \cref{eq:u*w*}--\cref{eq:V*}, along with $u^* \subseteq \uu^0(w^*)$.

\item   
Finally, we establish necessity of the three conditions. As shown in part 1, the optimality conditions of $\min_u\max_w V$ are equivalent to the stationary conditions \cref{eq:u*w*}, so a solution to $\min_u\max_w V$ exists only if $d \in R(M)$, and by part 2 the same is true of $\max_w \min_u V$.  Note that it is \textit{not} the case that $d \notin R(M)$ precludes a solution of the inner problem: for $M_{11}=0$, $M_{12}=0$, $M_{22}=-1$, and $d=(1,0)$, we have $d \notin R(M)$ and $\max_w V(u,w) = u$ exists for every $u$, while it is the outer minimization that has no solution.  If $M_{11}$ is not positive semidefinite, there is no solution to $\min_u V(u,w)$, and if $M_{22}$ is not negative semidefinite, there is no solution to $\max_w V(u,w)$, and hence strong duality does not hold. Therefore the conditions $d \in R(M)$, $M_{11} \geq 0$, and $M_{22} \leq 0$ are necessary for strong duality.
For sufficiency, note that $M_{11}\geq 0$ and $M_{22}\leq 0$ imply $\tM_{11}\leq 0$ and $\tM_{22} \geq 0$, and therefore  by parts 1 and 2, both $\min_u\max_w V$ and $\max_w\min_u V$ have solutions.
Comparing the optimal values in parts 1 and 2 establishes the sufficiency of these conditions for strong duality.
\end{enumerate} 
%We have thus established all the claims of the proposition.
\end{proof}

 Applying \cref{prop:SPalt} to the following example with $M_{11}=M_{22}=0$ and $M_{12}=1$
\begin{equation*}
 M = \begin{bmatrix} 0 & 1 \\ 1 & 0 \end{bmatrix}  \qquad V(u,w) = uw + \begin{bmatrix}u \\ w \end{bmatrix}' d
\end{equation*} 
gives $(u^*,w^*) = -(d_2, d_1)$, $V(u^*,w^*) = -d_1d_2$, $\ow^0(u^*) = \bbR$, $\uu^0(w^*) = \bbR$.  Note that both functions $\ow^0(\cdot)$ and $\uu^0(\cdot)$ are defined at only a single point, $u^*$ and $w^*$, respectively. So in this degenerate  case, these functions are not even differentiable.

\paragraph{Lagrangian functions.}
The connections between constrained optimization problems via the use of Lagrange multipliers and game theory problems are useful \citep{rockafellar:1993}. 

\begin{quote}
For optimization problems of convex type, Lagrange 
multipliers take on a game-theoretic role that could hardly even have been imagined
before the creative insights of von Neumann [32], [33], in applying mathematics to
models of social and economic conflict.\\

--R.T. Rockafellar
\end{quote}
\nocite{vonneumann:morgenstern:1944}

Next we are interested in the Lagrangian function $L(\cdot): \bbR^{m+n+1} \rightarrow \bbR$
\begin{align*}
 L(u,w,\lambda) &\eqbyd V(u,w) - (1/2) \lambda (w'w -1 ) \\
&= (1/2) \begin{bmatrix} u \\ w \end{bmatrix}'
\begin{bmatrix} M_{11} & M_{12} \\ M'_{12} & M_{22} \end{bmatrix}
\begin{bmatrix} u \\ w \end{bmatrix}
 + \begin{bmatrix} u \\ w \end{bmatrix}' d
 - (1/2) \lambda (w'w -1) \\
&= (1/2) \begin{bmatrix} u \\ w \end{bmatrix}'
\begin{bmatrix} M_{11} & M_{12} \\ M'_{12} & M_{22}-\lambda I \end{bmatrix}
\begin{bmatrix} u \\ w \end{bmatrix} 
 + \begin{bmatrix} u \\ w \end{bmatrix}' d
+ \lambda/2\\
&= (1/2) \begin{bmatrix} u \\ w \end{bmatrix}'
M(\lambda) \begin{bmatrix} u \\ w \end{bmatrix}
 + \begin{bmatrix} u \\ w \end{bmatrix}' d
 + \lambda/2
\end{align*} 
with 
\begin{equation*}
M(\lambda) \eqbyd 
\begin{bmatrix} M_{11} & M_{12} \\ M'_{12} & M_{22}-\lambda I \end{bmatrix}\\
\end{equation*} 
and $M(0) \in \bbR^{(m+n)\times (m+n)} \geq 0$, and $M_{11} \in \bbR^{m\times m}$, $M_{12} \in
\bbR^{m\times n}$, $M_{22} \in \bbR^{n\times n}$.

Note that from \cref{prop:psd}, both $M_{11}\geq 0$ and $M_{22}\geq 0$ as well, so that $\max_w$ is not bounded unless $\lambda$ is large enough to make $M_{22}-\lambda I \leq 0$. The Schur complements of $M_{11}$ and $M_{22}-\lambda I$ are useful for expressing the solution. 
\begin{align*}
 \tM_{11}(\lambda) &\eqbyd (M_{22}-\lambda I) - M_{12}'M_{11}^+ M_{12} = \tM_{11} - \lambda I\\
\tM_{22}(\lambda) &\eqbyd M_{11}-M_{12}(M_{22}-\lambda I)^+ M_{12}'
\end{align*}
Note that both Schur complements depend on the parameter $\lambda$.

\begin{subtheorem}{thm}
\setcounter{theorem}{14}
\setcounter{parentnumber}{\value{theorem}-1}
\end{subtheorem} 
\begin{subtheorem}{thm}
\begin{thm}[Alternative to 13 and 14]
\label{prop:sdparalt}

Consider the Lagrangian function $L(\cdot): \bbR^{m+n+1} \rightarrow \bbR$
\begin{align*}
L(u,w,\lambda) &\eqbyd (1/2) \begin{bmatrix} u \\ w \end{bmatrix}'
M(\lambda) \begin{bmatrix} u \\ w \end{bmatrix}  +
 \begin{bmatrix} u \\ w \end{bmatrix}' d + \lambda/2
\end{align*}
with $M_{11}>0$, $M(0) \geq 0$, and the two problems
\begin{equation}
\min_u \max_w L(u,w,\lambda) \qquad \max_w \min_u L(u,w,\lambda)
\label{eq:towprobs}
\end{equation} 

We characterize existence of solutions as a function of (decreasing) parameter $\lambda$.
\begin{enumerate}
\item For $\lambda > \norm{M_{22}}$: Solutions to both problems exist for all $d \in \bbR^{m+n}$.

\item  For $\lambda = \norm{M_{22}}$, we have the following two cases:
\begin{enumerate}
 \item For $d \in R(M(\norm{M_{22}}))$: The solutions to both problems exist.
 \item For $d \notin R(M(\norm{M_{22}}))$: Neither problem has a solution.
\end{enumerate} 

If $M(0)$ is such that $\normf{\tM_{11}} < \norm{M_{22}}$, then we have the following cases.
\item \label{case:int}
 For $\normf{\tM_{11}} < \lambda < \norm{M_{22}}$: Only the solution to the $\max_w \min_u L$ problem exists, and it exists for all $d \in \bbR^{m+n}$.
\item \label{case:edge}
 For $\lambda = \normf{\tM_{11}}$, we have the following two cases:
\begin{enumerate}
 \item For $d \in R(M(\normf{\tM_{11}}))$: Only the solution to the $\max_w \min_u L$ problem exists.
 \item For $d \notin R(M(\normf{\tM_{11}}))$: Neither problem has a solution.
\end{enumerate} 

\item For $\lambda < \normf{\tM_{11}}$:  Neither problem has a solution. 
\end{enumerate} 
If $M(0)$ is such that $\normf{\tM_{11}} = \norm{M_{22}}$, then cases \ref{case:int} and \ref{case:edge} do not arise.
    
For $d \in R(M(\lambda))$ denote the stationary points $(u^*(\lambda), w^*(\lambda))$ by
\begin{equation*}
\begin{bmatrix}u^* \\ w^* \end{bmatrix}(\lambda) \in -M^+(\lambda) d + N(M(\lambda))
\end{equation*} 
When solutions to the respective problems exist, we have that 
\begin{gather*}
u^*(\lambda) = \arg \min_u \max_w L(u,w,\lambda) \qquad w^*(\lambda) = \arg \max_w \min_u L(u,w,\lambda) \\
L^0(\lambda) = -(1/2)  d' M^+(\lambda) d + \lambda/2 
\end{gather*} 
and the inner optimizations, $\uu^0(w^*(\lambda))$ and $\ow^0(u^*(\lambda))$, are given by all solutions  to
% \begin{alignat*}{6}
% &M_{11} \; &&\uu^0(w^*) \; &&+ \; &&M_{12}             &&w^*        &&= - d_1\\
% &M_{12}'  \; &&u^*       \;  &&+ \; &&(M_{22}-\lambda I) \; &&\ow^0(u^*) &&=  -d_2
% \end{alignat*} 
\begin{equation*}
 \begin{array}{lccccl}
M_{11}  &\uu^0(w^*)  &+  &M_{12}             &w^*        &= - d_1\\
M_{12}'  &u^*        &+  &(M_{22}-\lambda I) &\ow^0(u^*) &=  -d_2
\end{array} 
\end{equation*} 
or, after solving,
% \begin{alignat*}{3}
% &\uu^0(w^*) = &&-M_{11}^+ (M_{12}w^* + d_1) &&+ N(M_{11})\\
% &\ow^0(u^*) = &&-(M_{22}-\lambda I)^+ (M_{12}'u^* + d_2) &&+ N(M_{22}-\lambda I)
% \end{alignat*}
\begin{equation*}
 \begin{array}{lcccl}
\uu^0(w^*) &= &-M_{11}^+ (M_{12}w^* + d_1) &+ &N(M_{11})\\
\ow^0(u^*) &= &-(M_{22}-\lambda I)^+ (M_{12}'u^* + d_2) &+ &N(M_{22}-\lambda I)
\end{array} 
\end{equation*} 
 \end{thm} 
\end{subtheorem}

\begin{proof}
First note that for symmetric $M(\lambda)$, $M_{11}>0$ and $M_{22}-\lambda I < 0$ imply that $M(\lambda)$ is full rank, i.e., just set $M(\lambda)x=0$ and show that $x=0$.  We also use the following fact: if $A$ and $B$ are compatible, square matrices and $A$ is full rank, then $B$ is full rank if and only if $AB$ is full rank.

Next we apply \cref{prop:SPalt} for the different values of $\lambda$.
\begin{enumerate}
\item For $\lambda > \norm{M_{22}}$ the conditions for parts 1 and 2 of \cref{prop:SPalt} are satisfied so both solutions exist for $d \in R(M(\lambda))$, and since $M_{11} > 0$ and $M_{22} - \lambda I <0$, $M(\lambda)$ is full rank, and $R(M(\lambda)) = \bbR^{m+n}$.

\item For $\lambda = \norm{M_{22}}$, conditions for 1 and 2 of \cref{prop:SPalt} still hold, so both solutions exist for $d \in R(M(\norm{M_{22}}))$, but $M(\norm{M_{22}})$ is no longer full rank giving cases 2(a) and 2(b).

\item For $\normf{\tM_{11}} < \lambda < \norm{M_{22}}$, $M_{22} - \lambda I$ is no longer negative semidefinite, and from \cref{prop:SPalt} the solution to $\min_u\max_w L$ no longer exists.
But $M_{11}>0$ and $\tM_{11}(\lambda)< 0$ so the solution to $\max_w\min_u$ still exists for all $d \in R(M(\lambda))$.  Also note that we can multiply $M(\lambda)$ by a full rank matrix to obtain
\begin{equation*}
\begin{bmatrix} I & 0 \\ -M_{12}'M_{11}^{-1} & I \end{bmatrix} M(\lambda) =
\begin{bmatrix} M_{11} & M_{12} \\ 0 & \tM_{11}(\lambda) \end{bmatrix}
\end{equation*}
Since the last matrix is full rank, and the first matrix is full rank, we know that $M(\lambda)$ is full rank, and therefore the solution to $\max_w\min_u$ exists for all $d \in \bbR^{m+n}$.

\item For $\normf{\tM_{11}}=\lambda$, $\tM_{11}(\lambda)$ is still negative semidefinite, so the solution to $\max_w\min_u L$ still exists for $d \in R(M(\lambda))$, but $M(\lambda)$ is no longer full rank giving cases 4(a) and 4(b).

\item For  $\lambda < \normf{\tM_{11}}$, $\tM_{11}(\lambda)$ is no longer negative semidefinite and the solution to $\max_w\min_u L$ no longer exists, so neither problem has a solution.
\end{enumerate}
Applying the formulas in \cref{prop:SPalt} for the optimal solutions and cost completes the proof. 
\end{proof}

\begin{figure}
\centerline{\includegraphics[width=0.8\textwidth]{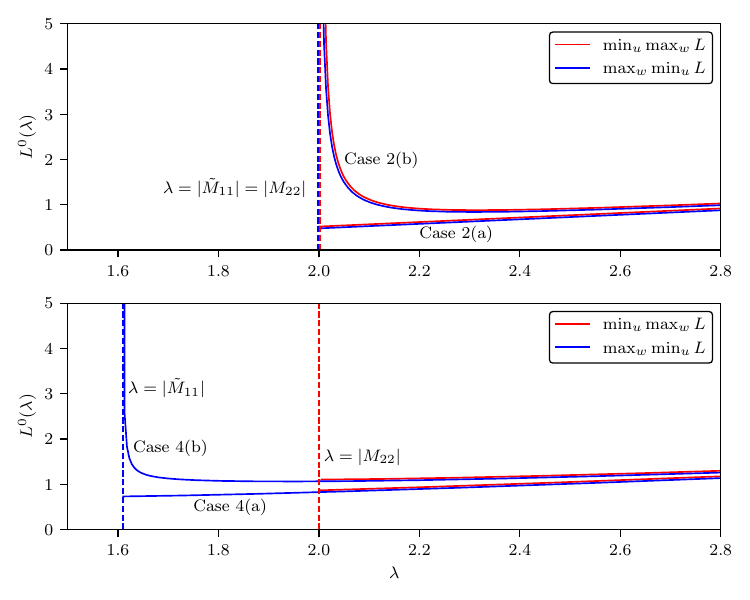}}
\caption{The optimal value function $L^0(\lambda)$ for $\min_u\max_wL$ and $\max_w\min_uL$ versus parameter $\lambda$. % Strong duality holds only when $\lambda \geq \norm{M_{22}}$. For $\lambda < \norm{M_{22}}$, $\min_u \max_w L(u,w,\lambda) = +\infty$. 
% For $\lambda < \normf{\tM_{11}}$, $\max_w \min_u L(u,w,\lambda) = +\infty$.
Top: $\normf{\tM_{11}} = \norm{M_{22}}$ showing cases 2(a) and 2(b).  Bottom: $\normf{\tM_{11}} < \norm{M_{22}}$ showing cases 4(a) and 4(b). } 
\label{fig:sdparalt}
\end{figure}

Figure \ref{fig:sdparalt} illustrates the different possible cases for \cref{prop:sdparalt}.
Note that in \cref{prop:sdparalt}, for cases where solutions to both problems exist, strong duality holds. For cases where only the solution to $\max_w \min_u L$  exists, the duality gap is infinite. 
For strong duality to hold for all $\lambda$ such that either problem has a bounded solution requires that $\norm{M_{22}} = \normf{\tM_{11}}$.  The following example, with $m=1$ and $n=2$, shows that it is not \textit{necessary} for $M_{12}'M_{11}^+M_{12} = 0$ for this condition to hold.
\begin{equation*}
M_{11} = 1  \qquad 
M_{12} = \begin{bmatrix} 0 & 1 \end{bmatrix}   \qquad 
M_{22} = \begin{bmatrix} 1 & 0\\ 0 & 1 \end{bmatrix}
\end{equation*}
\begin{equation*}
M_{12}'M_{11}^{-1}M_{12} = \begin{bmatrix} 0 & 0 \\ 0 & 1 \end{bmatrix} \qquad 
\tM_{11} = M_{22} - M_{12}' M_{11}^{-1}M_{12} = \begin{bmatrix} 1 & 0 \\ 0 & 0 \end{bmatrix}
\end{equation*}
This example satisfies the assumptions of \cref{prop:sdparalt}: $M_{11}>0$, and $M(0) \geq 0$ by \cref{prop:psd} since $M_{11} > 0$ and $\tM_{11} \geq 0$.  We have that $\norm{M_{22}}=1$ and $\normf{\tM_{11}} = 1$, so the norms are equal but $M_{12}'M_{11}^{-1}  M_{12} \neq 0$.

An unstated implication of \cref{prop:sdparalt} is that case 2(b) is incompatible with either case 4(a) or case 4(b), as illustrated in \cref{fig:sdparalt}. In other words, if $M(0)$ is such that $\normf{\tM_{11}} < \norm{M_{22}}$, then $M(\norm{M_{22}})$ is full rank, its range is therefore $\bbR^{m+n}$, and there is no $d$ satisfying case 2(b).  To establish this implication, first note from item 3 of the proof of \cref{prop:sdparalt} that $M(\lambda)$ is full rank if and only if both $M_{11}$ and $\tM_{11}(\lambda)$ are full rank. Matrix $M_{11}$ is full rank since $M_{11}>0$ by assumption. From its definition, $\tM_{11}(\lambda) < 0$, and therefore full rank, for $\lambda > \normf{\tM_{11}}$, which gives that $M(\lambda)$ is full rank for all $\lambda > \normf{\tM_{11}}$. Since $\tM_{11} \leq M_{22}$ implies $\normf{\tM_{11}} \leq \norm{M_{22}}$, we conclude that $M(\norm{M_{22}})$ is full rank unless $\norm{M_{22}} = \normf{\tM_{11}}$.

\paragraph{Constrained quadratic optimization}

\begin{quote}
A mysterious piece of information has been uncovered. In our innocence we
thought we were engaged straightforwardly in solving a single problem (P). But we
find we’ve assumed the role of Player 1 in a certain game in which we have an adversary,
Player 2, whose interests are diametrically opposed to ours!\\

--R.T. Rockafellar
\end{quote}

We next consider \textit{maximization} of a \textit{convex} function so that a constraint is required for even existence of a solution.  We establish the following result.

\begin{figure}
\centerline{\includegraphics[width=0.85\textwidth]{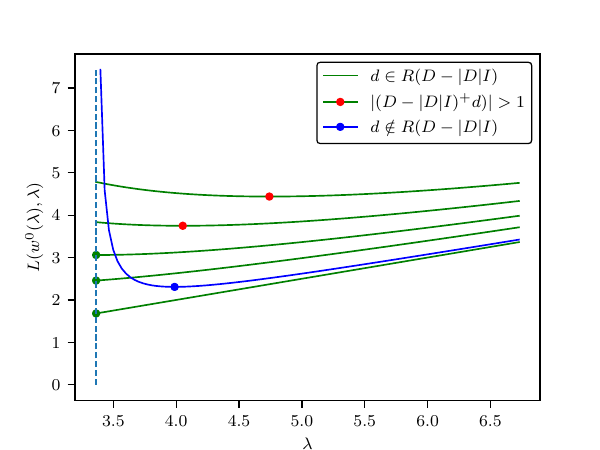}}
\caption{$L(w^0(\lambda), \lambda)$ versus $\lambda$ for the same $D$ but different $d$.
Green lines: for $d \in R(D-\norm{D}I)$, $L$ is bounded for all $\lambda\geq \norm{D}$.
Green dots: for $\norm{(D-\norm{D}I)^+d} \leq 1$, the optimum is on the boundary and $\lambda_P= \norm{D}$.
Red dots: for $\norm{(D-\norm{D}I)^+d} > 1$, the optimum is in the interior and $\lambda_P> \norm{D}$.
Blue line and dot: for $d \notin R(D-\norm{D}I)$, $L$ is unbounded at $\lambda = \norm{D}$, and
the optimum is in the interior, and $\lambda_P > \norm{D}$.}
\label{fig:conquad}
\end{figure}

\begin{proposition}[Constrained quadratic optimization]
\label{prop:conquad}

Define the \textit{convex} quadratic function, $V(\cdot): \bbR^{n} \rightarrow \bbR$ and compact constraint set $W$
\begin{equation*}
V(w) \eqbyd \half w'Dw + w'd \qquad 
W \eqbyd \{ w \mid w'w = 1\}
\end{equation*}
with $D \in \bbR^{n\times n} \geq 0$. Consider the constrained maximization problem
\begin{equation}
\max_{w \in W} V(w)
\label{eq:maxcon}
\end{equation}
Define the Lagrangian function
\begin{equation*}
L(w, \lambda) = V(w) - \half \lambda (w'w-1)
%\label{eq:Ldef}
\end{equation*}
and the (unconstrained) Lagrangian problem
\begin{equation}
\max_{w} \min_{\lambda}  L(w, \lambda)
\label{eq:maxmincon}
\end{equation}
and the (unconstrained) dual Lagrangian problem
\begin{equation}
\min_{\lambda} \max_{w} L(w, \lambda)
\label{eq:minmaxcon}
\end{equation}

\begin{enumerate}
\item  Solutions to all three problems \eqref{eq:maxcon}, \eqref{eq:maxmincon}, and \eqref{eq:minmaxcon} exist for all $D \geq 0$ and $d \in \bbR^n$ with optimal value
\begin{equation*}
V^0 = L^0 = - (1/2) d'(D-\lambda_P I)^+ d + \lambda_P/2
\end{equation*}
where
\begin{equation}
\lambda_P := \; \text{the largest real eigenvalue of $P$}
\qquad 
P \eqbyd \begin{bmatrix} D & I \\ dd' & D \end{bmatrix}
\label{eq:Pdef}
\end{equation}

\item Problems \eqref{eq:maxmincon} and \eqref{eq:minmaxcon} satisfy strong duality,
and the function $L(w, \lambda)$ has saddle points (sets) $(w^*, \lambda^*)$ given by
\begin{align*}
w^* &= \begin{cases} \bigg( -(D-\lambda_P I)^+ d + N(D-\lambda_P I) \bigg) \cap W, \quad & \lambda_P = \norm{D} \\
                     -(D-\lambda_P I)^{-1} d, \qquad & \lambda_P > \norm{D}
\end{cases} \\
\lambda^* &= \lambda_P
\end{align*}

\item The optimizer of \eqref{eq:maxcon} is given by $w^0 = w^*$. 

\item 
The optimizer of \eqref{eq:maxmincon} is given by 
\begin{equation*}
w^0 = w^* \qquad 
\ulambda^0(w^0) = \bbR
\end{equation*}

\item  The optimizer of \eqref{eq:minmaxcon} is given by
\begin{align*}
\ow^0(\lambda^0) &= \begin{cases} -(D-\lambda_PI)^+ d + N(D-\lambda_P I), \quad & \lambda_P = \norm{D} \\
                     -(D-\lambda_PI)^{-1} d, \qquad & \lambda_P > \norm{D}
\end{cases} \\
\lambda^0 &= \lambda_P 
\end{align*}

\item Additionally $\lambda_P = \norm{D}$ if and only if (i) $d \in R( D - \norm{D} I)$ and (ii) $\norm{(D-\norm{D}I)^+ d} \leq 1$. If (i) or (ii) do not hold, then $\lambda_P > \norm{D}$ and $\norm{(D-\lambda_P I)^{-1}d} = 1$.
\end{enumerate}
\end{proposition}
Figure \ref{fig:conquad} shows the possible behaviors. The green lines show the case $d \in R(D-\norm{D}I)$,  for different $\norm{d}$.  For $d=0$ (bottom green line), $\lambda_P=\norm{D}$ and the optimum is on the boundary. Increasing $\norm{d}$ eventually produces a zero derivative at $\lambda=\norm{D}$ (third green line from bottom). Further increasing $\norm{d}$ makes the derivative at $\lambda=\norm{D}$ negative (top two green lines), and $\lambda_P >  \norm{D}$ (red dots), and the optimum moves to the interior.
The blue line shows the case $d \notin R(D-\norm{D}I)$. $L$ is unbounded at $\lambda = \norm{D}$, $\lambda_P > \norm{D}$ (blue dot) and the optimum is again in the interior. 
 
To organize the proof of this proposition, we treat the Lagrangian, dual Lagrangian, and saddle-point problems in separate lemmas, and then combine them.  We start with the dual Lagrangian minmax problem. We shall find that all of the information about $\lambda_P$ emerges from this problem. 

\begin{lemma}[Dual Lagrangian of constrained quadratic optimization]
\label{lem:dLquad}
Consider the dual Lagrangian problem
\begin{equation}
\min_\lambda \max_w L(w,\lambda) \qquad L(w, \lambda) \eqbyd (1/2) w'Dw + w'd -(1/2)\lambda (w'w-1)
\label{eq:dL}
\end{equation}
with $D\geq 0$.   We have the following results.
\begin{enumerate}
\item 
This problem is equivalent to 
\begin{equation*}
\min_{\lambda \geq \norm{D}} \max_{w} L(w, \lambda)
\end{equation*}
\item The solution exists for all $D$ and $d$ and has optimal value
\begin{equation}
L^0 = -(1/2) d' (D - \lambda_P I)^{+} d + \lambda_P/2
\label{eq:dLsol}
\end{equation}
where $\lambda_P$ and matrix $P \in \bbR^{2n \times 2n}$ are defined as
\begin{equation}
\lambda_P \eqbyd \; \text{the largest real eigenvalue of $P$}
\qquad 
P \eqbyd \begin{bmatrix} D & I \\ dd' & D \end{bmatrix}
%\label{eq:Pdef_rep}
\end{equation}
\item The optimal $\lambda$ and $w^0(\lambda)$ are given by
\begin{align*}
\lambda^0 &= \lambda_P \in [\norm{D}, \infty) \\
w^0(\lambda^0) &= \begin{cases}
-(D-\norm{D} I)^+ d + N(D-\norm{D} I), \quad & \lambda_P = \norm{D} \\
-(D-\lambda_P I)^{-1} d, \quad & \lambda_P \in (\norm{D}, \infty)
\end{cases}
\end{align*}

\item We have that $\lambda_P = \norm{D}$ if and only if (i) $d \in R(D - \norm{D}I)$, and (ii) $\norm{(D-\norm{D}I)^+d} \leq 1$. Otherwise $\lambda_P > \norm{D}$.  If (i) is violated, $L(w^0(\lambda), \lambda) = +\infty$ at $\lambda=\norm{D}$. If (i) holds but (ii) is violated, then $L(w^0(\lambda, \lambda)$ is finite at $\lambda=\norm{D}$, but $(d/d\lambda) L(w^0(\lambda), \lambda) < 0$ at $\lambda=\norm{D}$.
\end{enumerate}

\end{lemma}

\begin{proof}
To establish statement 1 in the lemma, note that 
if $\lambda < \norm{D}$, then $D-\lambda I$ has the positive eigenvalue $\norm{D}-\lambda$, so $L(\cdot,\lambda)$ is unbounded above along the corresponding eigenvector and $\max_w L(w, \lambda) = +\infty$.  So adding the constraint $\lambda \geq \norm{D}$ to the outer minimization does not alter the solution. 

To establish statements 2--4 in the lemma, we make use of the SVD of matrix $D = UMU'$, which we partition as 
\begin{equation*}
D = \begin{bmatrix}U_1 & U_2 \end{bmatrix} \begin{bmatrix} \norm{D}I_p &  \\ & M_2 \end{bmatrix}
\begin{bmatrix}U_1' \\ U_2' \end{bmatrix}
\end{equation*}
where $p$ is the  multiplicity of the largest eigenvalue of $D$, $1 \leq p  \leq n$.
Also denote $y = U'd, y_1=U_1'd, y_2 = U_2'd$.

We break the problem into two cases according to whether the derivative of $L(w^0(\lambda),\lambda)$ vanishes on the open interval $(\norm{D}, \infty)$; note that we cannot split on the location of $\lambda_P$, because the existence of a real eigenvalue of $P$ is one of the facts to be established.
\begin{enumerate}
\item Case $(d/d\lambda)L(w^0(\lambda),\lambda) = 0$ for some $\lambda \in (\norm{D}, \infty)$.

For $\lambda \in (\norm{D}, \infty)$, we have from \cref{prop:min} that $w^0(\lambda)=-(D-\lambda I)^{-1}d$ and $L(w^0(\lambda),\lambda) = -(1/2) d'(D-\lambda I)^{-1}d + \lambda/2$ .  $L(w^0(\lambda), \lambda)$ is
differentiable, and taking two derivatives gives
\begin{align*}
\frac{dL}{d\lambda} &= (1/2)( 1 - d'(D-\lambda I)^{-2} d) \\
\frac{d^2L}{d\lambda^2} &= - d'(D - \lambda I)^{-3} d
\end{align*}
Setting the first derivative to zero yields
\begin{align}
0 &= 1 - d'(D - \lambda I)^{-2} d \label{eq:line1}\\
&= \det (1 - d'(D - \lambda I)^{-2} d) \notag \\
&= \det (I - dd'(D - \lambda I)^{-2}) \notag \\
&= \det \left( (D-\lambda I) - dd'(D - \lambda I)^{-1} \right)
\det(D-\lambda I)^{-1} \notag
\end{align}
where we have used the fact that $\det(I+AB)=\det(I+BA)$.
Since $\det(D-\lambda I) \neq 0$, we can multiply both sides of the last equality by $\det (D-\lambda I)$  
to obtain
\begin{align*}
0 &= \det \left( (D-\lambda I) - dd'(D - \lambda I)^{-1} \right) \det (D - \lambda I) \\
&= \det \left( \begin{bmatrix} D - \lambda I & I \\ dd' & D - \lambda
    I \end{bmatrix} \right) = \det (P - \lambda I)
\end{align*}
where we have used the partitioned determinant formula, which is valid since
$D-\lambda I$ is nonsingular for $\lambda > \norm{D}$.  
Therefore the first derivative of $L$ vanishes in the interval $(\norm{D}, \infty)$ 
if and only if there exists a real-valued eigenvalue of $P$  in this
interval. Also, we have from \cref{eq:line1} that
\begin{equation*}
1 = d'(D - \lambda I)^{-2} d = d' U (M - \lambda I)^{-2} U' d =
y'(M-\lambda I)^{-2} y
\end{equation*}
with $y \eqbyd U'd$.  So we conclude that $y \neq 0$ for this case. 
Examining the second derivative, we have
\begin{equation*}
\frac{d^2L}{d\lambda^2} = d'(\lambda I - D)^{-3} d
= y'(\lambda I - M)^{-3} y
\end{equation*}
Note that since $\lambda > \norm{D}$, $(\lambda I - M)^{-3}>0$, and since $y \neq
0$, we have that $d^2L/d\lambda^2> 0$ on $(\norm{D}, \infty)$ and therefore
$L(w^0(\lambda), \lambda)$ is strictly convex on this interval.
Therefore the minimizer of $L$ is unique and the first derivative is zero at the solution. We also know 
that there is only one real eigenvalue of $P$ in this interval due to the uniqueness of the optimal solution. 
Therefore we have established that $\lambda^0 = \lambda_P \in (\norm{D}, \infty)$
is the optimal solution. Substituting this solution into $L$ gives
\begin{equation*}
L^0 = L(w^0(\lambda^0), \lambda^0) = -(1/2) d'(D-\lambda_PI)^{-1}d + \lambda_P/2
\end{equation*}
verifying that \eqref{eq:dLsol} holds for the first case.

\item Case $(d/d\lambda)L(w^0(\lambda),\lambda) \neq 0$ for all $\lambda \in (\norm{D}, \infty)$.
In this case, we first show that the $\lambda^0 = \norm{D}$.  
Using the SVD, we have for $\lambda \in (\norm{D}, \infty)$
\begin{align*}
L(w^0(\lambda), \lambda) &= -(1/2) d'(D-\lambda I)^{-1}d + \lambda/2  \\
&= -(1/2) \begin{bmatrix}y_1' & y_2' \end{bmatrix} 
\begin{bmatrix} \frac{1}{\norm{D} -\lambda}I_p & \\ & (M_2-\lambda I)^{-1} \end{bmatrix} 
\begin{bmatrix}y_1 \\ y_2 \end{bmatrix} + \lambda/2
\end{align*}
From this expression for $L$, note that $y_1=U_1'd$ must be zero for this case, or $\lim_{\lambda \rightarrow \norm{D}^+} L(w^0(\lambda), \lambda) = +\infty$, which is a contradiction since $\lim_{\lambda \rightarrow +\infty}L(w^0(\lambda), \lambda) = +\infty$ as well, and $L$ is a smooth function on the interval $(\norm{D}, \infty)$, so it must have a minimum on that interval (zero derivative), but by assumption it does not have a zero derivative on that interval. 
Note that $y_1=U_1'd =0$ is equivalent to $d \in R(D-\norm{D}I)$, which can be seen from the SVD of $D-\norm{D}I$
\begin{equation*}
D - \norm{D} I = \begin{bmatrix}U_1 & U_2 \end{bmatrix} \begin{bmatrix} 0 &  \\ & M_2-\norm{D}I \end{bmatrix}
\begin{bmatrix}U_1' \\ U_2' \end{bmatrix}
\end{equation*}
so the columns of $U_1$ are a basis for $N(D-\norm{D}I)$ and $d$ is orthogonal to the columns of $U_1$ so 
$d \in R(D-\norm{D}I)$. 
Substituting $y_1=0$, into the expression for $L(w^0(\lambda), \lambda)$ gives
\begin{equation}
L(w^0(\lambda), \lambda) = -(1/2)d' (D-\lambda I)^+ d + \lambda/2, \qquad \lambda \geq \norm{D}, \; d \in R(D-\norm{D}I)
\label{eq:Lboundary}
\end{equation}
and $L(w^0(\lambda), \lambda)$ is smooth on the interval \textit{including} the left boundary, $[\norm{D}, \infty)$, and the optimizer must be on the boundary, $\lambda^0 = \norm{D}$.
For this value of $\lambda$, the inner maximization over $w$ gives from \cref{prop:min}
\begin{equation*}
w^0(\lambda^0) = -(D-\norm{D}I)^+ d + N(D-\norm{D}I)
\end{equation*}
and evaluating $L^0$ gives
\begin{align}
L(w^0(\lambda^0), \lambda^0) &= -(1/2)  y_2' (M_2-\norm{D}I)^{-1}y_2 + \norm{D}/2  \notag \\
&= -(1/2) d' (D-\norm{D}I)^+ d +  \norm{D}/2 \qquad 
\label{eq:opt2ndcase}
\end{align}
verifying \cref{eq:dLsol} for this case.

Taking the derivative of \cref{eq:Lboundary} and evaluating at $\lambda = \norm{D}$ gives
\begin{equation*}
(d/d \lambda) L(w^0(\lambda), \lambda) = (1/2) (1- d'((D-\norm{D}I)^+)^2 d) = (1/2)(1-\norm{(D-\norm{D}I)^+d}^2) 
\end{equation*}
which is non-negative if and only if 
$\norm{(D-\norm{D}I)^+d} \leq 1$.  Otherwise the derivative at the boundary is negative and the optimal $\lambda$ is in the interval $(\norm{D}, \infty)$, which is the previous case.  Therefore $\lambda^0 = \norm{D}$ if and only if
\begin{equation*}
d \in R(D - \norm{D}I), \quad \norm{(D-\norm{D}I)^+d} \leq 1
\end{equation*}
Next we show that $\norm{D}$ is an eigenvalue of $P$ in this case. Factoring $P - \lambda I$ gives
\begin{align*}
P - \lambda I 
&= \begin{bmatrix} U & \\ & U \end{bmatrix} 
\begin{bmatrix} U'D U -\lambda I & I \\ U'd d'U & U'DU - \lambda I \end{bmatrix} 
\begin{bmatrix} U' & \\ & U' \end{bmatrix} \\
&= \begin{bmatrix} U & \\ & U \end{bmatrix} 
\begin{bmatrix}
(\norm{D} - \lambda)I_p & & I &  \\
 & M_2 - \lambda I &  & I \\
y_1y_1' & y_1 y_2' & (\norm{D} - \lambda)I_p & \\
y_2y_1' & y_2 y_2' & & M_2 - \lambda I 
\end{bmatrix}
\begin{bmatrix} U' & \\ & U' \end{bmatrix}
\end{align*}
Since the leading and trailing matrices are inverses of each other, we have a similarity
transformation, and the eigenvalues of the inner matrix are the eigenvalues of $P$.  Setting $y_1=0$ in the inner matrix and setting $\lambda = \norm{D}$ gives a zero third block row of the inner matrix, and it is singular. Therefore $\lambda=\norm{D}$ is an eigenvalue of $P$.  Since there are no real eigenvalues of $P$ in $(\norm{D}, \infty)$, we have that $\norm{D}$ is the \textit{largest} real eigenvalue of $P$ for this case, and we have established that $\lambda^0 = \lambda_P$ also for this case.
\end{enumerate}
Summarizing, we have broken the problem into two cases.  In the first case we have shown that $\lambda^0 = \lambda_P > \norm{D}$, and $(d/d\lambda) L(w^0(\lambda), \lambda)$ is zero at $\lambda = \lambda_P$. In this case there is only one real eigenvalue of $P$ in $(\norm{D}, \infty)$.

In the second case, we have that $\lambda^0=\lambda_P = \norm{D}$, the boundary of the feasible set. We have also shown that $d \in R(D - \norm{D}I)$, and $\norm{(D-\norm{D}I)^+d} \leq 1$ for this case.  If $d \notin R(D-\norm{D}I)$, then $L(w^0(\lambda), \lambda)$ is $+\infty$ at $\lambda=\norm{D}$, which is in the first case. If $d \in R(D-\norm{D}I)$, but $\norm{(D-\norm{D}I)^+d} > 1$, then $L(w^0(\lambda), \lambda)$ is finite at $\lambda=\norm{D}$, but the derivative is negative, which is again in the first case.  Thus we have established statements 2--4 in the lemma and the proof is complete.
\end{proof}

Next we turn to the saddle points. 
\begin{lemma}[Saddle points of the Lagrangian of constrained quadratic optimization]
\label{lem:SPlag}
The following $(w^*, \lambda^*)$ are saddle points of $L(w, \lambda) \eqbyd (1/2) w'Dw + w'd -(1/2)\lambda (w'w-1)$. 
\begin{align*}
w^* &= \begin{cases}
\bigg( -(D-\norm{D} I)^{+} d + N(D-\norm{D}I) \bigg) \cap W \quad & \lambda_P = \norm{D}\\
-(D-\lambda_P I)^{-1} d, \quad & \lambda_P > \norm{D} 
\end{cases}  \\
\lambda^* &= \lambda_P 
\end{align*}
\end{lemma}

\begin{proof}
From the definition of a saddle point we need to establish the inequalities
\begin{equation*}
L(w, \lambda^*) \leq L(w^*, \lambda^*) \leq L(w^*,\lambda)
\end{equation*}
hold for all $w \in \bbR^n$ and  $\lambda \in \bbR$.

We first check that $w^*$ is nonempty and satisfies $(w^*)'w^*=1$ in both cases, which is used below.  For $\lambda_P > \norm{D}$, evaluating \cref{eq:line1} at $\lambda = \lambda_P$ gives $d'(D-\lambda_PI)^{-2}d = 1$, which is $\norm{(D-\lambda_PI)^{-1}d} = 1$, so $w^*$ is a single point of $W$.  For $\lambda_P = \norm{D}$, the vector $-(D-\norm{D}I)^+d$ lies in $R(D-\norm{D}I)$ and is therefore orthogonal to $N(D-\norm{D}I)$, so for any $q \in N(D-\norm{D}I)$
\begin{equation*}
\norm{-(D-\norm{D}I)^+d + q}^2 = \norm{(D-\norm{D}I)^+d}^2 + \norm{q}^2
\end{equation*}
Since $D \geq 0$, $\norm{D}$ is an eigenvalue of $D$ and $N(D-\norm{D}I)$ is nontrivial, so a $q$ with $\norm{q}^2 = 1 - \norm{(D-\norm{D}I)^+d}^2$ exists if and only if $\norm{(D-\norm{D}I)^+d} \leq 1$, which is condition (ii) of statement 4 of \cref{lem:dLquad} and therefore holds whenever $\lambda_P = \norm{D}$.  So the intersection with $W$ is nonempty and every element of $w^*$ satisfies $(w^*)'w^* = 1$.

Taking the second inequality first, we have that $L(w^*, \lambda)= (1/2) (w^*)'Dw^* + (w^*)'d$
for all $\lambda$ since $(w^*)'w^* = 1$. Therefore $L(w^*, \lambda) = L(w^*,
\lambda^*)$ for all $\lambda$ and the second inequality is established with
equality.

Turning to the first inequality, we consider the two cases; (i) $\lambda^* =\lambda_P > \norm{D}$, and (ii) $\lambda^* =  \lambda_P = \norm{D}$ and $d \in R(D-\norm{D}I)$. For the first case we know that
\begin{equation*}
\max_w L(w, \lambda_P) = -(1/2) d'(D-\lambda_PI)^{-1}d + \lambda_P/2 = L(w^*,
\lambda_P) = L(w^*, \lambda^*)
\end{equation*}
where the maximization over $w$ is unconstrained, and the first equality follows from
\cref{prop:min}.  Since this equality holds  for the maximizer, we have that $L(w, \lambda^*) = L(w,
\lambda_P) \leq L(w^*, \lambda^*)$ for all $w \in \bbR^n$, where the first equality comes from the definition of $\lambda^*$.  Thus the first  inequality holds for the first case.  

Turning to the second case, we have similarly from \cref{prop:min}
\begin{equation*}
\max_w L(w, \norm{D}) = -(1/2) d'(D-\norm{D}I)^+d + \norm{D}/2= L(w^*, \norm{D}) = L(w^*, \lambda^*)
\end{equation*}
where again the maximization over $w$ is unconstrained, and we have $L(w, \lambda^*) = L(w, \norm{D})  \leq L(w^*, \lambda^*)$ for all $w \in \bbR^n$. Thus both cases satisfy the first inequality, both inequalities have been established,
and the result  is proven.
\end{proof}

Finally, we address the original constrained optimization of the concave quadratic function and its Lagrangian 
\begin{lemma}[Constrained quadratic optimization and its Lagrangian]
\label{lem:Lquad}
We are given the following: (i) a convex quadratic function and compact constraint set $W$ defined in \cref{prop:conquad}
\begin{equation*}
V(w) \eqbyd \half w'Dw + w'd \qquad 
W \eqbyd \{ w \mid w'w = 1\}
\end{equation*}
with $D \in \bbR^{n\times n} \geq 0$, (ii) the constrained maximization problem \cref{eq:maxcon}
\begin{equation*}
\max_{w \in W} V(w)
%\label{eq:maxcon}
\end{equation*}
with Lagrangian function
\begin{equation*}
L(w, \lambda) = V(w) - \half \lambda (w'w-1)
%\label{eq:Ldef}
\end{equation*}
and the (unconstrained) Lagrangian problem \cref{eq:maxmincon}
\begin{equation*}
\max_{w} \min_{\lambda}  L(w, \lambda)
%\label{eq:maxmincon}
\end{equation*}

\begin{enumerate}
\item 
Solutions to \eqref{eq:maxcon} and \eqref{eq:maxmincon} exist for all $D \geq 0$ and $d \in \bbR^n$ and achieve the same optimal value $V^0  = L^0 = -(1/2)d'(D-\lambda_PI)^+ d + \lambda_P/2$. 

\item The optimizer of \cref{eq:maxcon}, denoted $w^0$,  is given by $w^0=w^*$ where $w^*$ the saddle-point solution set from \cref{lem:SPlag}. 

\item The optimizer of \cref{eq:maxmincon}, denoted $(w_L^0, \ulambda(w_L^0))$ is given by $w_L^0=w^*$ and $\ulambda(w_L^0)=\bbR$. 
\end{enumerate}

\end{lemma}

\begin{proof}\mbox{ }

The solution to \eqref{eq:maxcon} exists since $V(\cdot)$ is continuous and $W$ is compact. 
The solution to \cref{eq:maxmincon} exists and satisfies strong duality with \eqref{eq:minmaxcon} 
due to the saddle-point theorem (\cref{prop:saddle}), which applies here as noted after its proof, and \cref{lem:SPlag}, so we have that
\begin{equation*}
L^0 = \max_w\min_\lambda L = \min_\lambda \max_w L = -(1/2) d' (D - \lambda_P I)^{+} d + \lambda_P/2
\end{equation*}
where the last equality follows by \eqref{eq:dLsol}. From the saddle-point theorem, we also have that
$w_L^0 = w^*$ for the optimizer of the Lagrangian. Since $w_L^0$ satisfies $(w_L^0)'w_L^0=1$, it follows that $\ulambda^0(w_L^0) = \bbR$.
Finally, we have that value $V^0=L^0$ and set $w^0=w_L^0$ by \cref{prop:conlag}, and the proof is complete. 
\end{proof}

The proofs of \cref{lem:dLquad}, \cref{lem:SPlag}, and \cref{lem:Lquad} have proven \cref{prop:conquad}. 

\paragraph{Discussion of \cref{prop:conquad}.}

The basic problem of constrained, nonconvex quadratic optimization has appeared in several fields. In the optimization literature it is known as the ``trust-region'' problem. 
\citeauthor{nocedal:wright:2006} discuss the numerical solution of the trust-region problem in the context of nonlinear programming \citep[p.69]{nocedal:wright:2006}. \citeauthor{boyd:vandenberghe:2004} establish strong duality of the Lagrangian and dual Lagrangian formulations of the problem \citep[Appendix B]{boyd:vandenberghe:2004}.
The problem also shows up in numerical linear algebra, where \citet{forsythe:golub:1965} characterize the stationary values of a second-degree polynomial on the unit sphere. 

The multiplier $\lambda_P$ defined in \eqref{eq:Pdef} has also appeared in the literature.  Eliminating $w$ between the stationarity condition $(D-\lambda I)w = -d$ and the constraint $w'w=1$ gives the secular equation $d'(D-\lambda I)^{-2}d = 1$, and setting $w_1 = (D-\lambda I)^{-2}d$, $w_2 = (D-\lambda I)w_1$ linearizes it as the $2n \times 2n$ eigenvalue problem
\begin{equation*}
\begin{bmatrix} D & -I \\ -dd' & D \end{bmatrix}
\begin{bmatrix} w_1 \\ w_2 \end{bmatrix} = \lambda
\begin{bmatrix} w_1 \\ w_2 \end{bmatrix}
\end{equation*}
which is similar to $P$ under $\diag(I,-I)$ and therefore has the same spectrum.  This is equation (34) of \citet{gander:golub:vonmatt:1989}, whose extremal eigenvalue solves the problem in a single step rather than by iteration on the secular equation.  That result drew little attention until \citet{adachi:iwata:nakatsukasa:takeda:2017}, who call it ``a relatively underappreciated fact,'' rederive it, and treat the computation of the $\lambda_P = \norm{D}$ case.  That case is part 6 of \cref{prop:conquad} and is known as the \textit{hard case} in the trust-region literature \citep{more:sorensen:1983}.

What appears to be completely new here is the assembly.  \cref{prop:conquad} collects the optimizers of the constrained problem, the Lagrangian, and the dual Lagrangian, together with the saddle points of $L(w,\lambda)$ and the strong duality relating them, all in terms of the single quantity $\lambda_P$; the trust-region literature states optimality conditions and algorithms rather than this minmax structure.
The authors would also like to acknowledge Robin Str{\"a}sser for his work on earlier versions of \cref{prop:conquad} \citep{mannini:straesser:rawlings:2024}.
 
\paragraph{Constrained minmax and maxmin.}

To compactly state the results in this section recall the definitions of the two functions
\begin{align}
M(\lambda) &\eqbyd \begin{bmatrix} M_{11} & M_{12} \\ M'_{12} & M_{22}-\lambda I \end{bmatrix} \label{eq:Mlam}\\
L(\lambda) &\eqbyd \begin{cases} - (1/2) d'M^+(\lambda)d + \lambda/2, \quad & d \in R(M(\lambda)) \\
+ \infty, \quad & d \notin R(M(\lambda)) \end{cases}
 \label{eq:Llam}
\end{align}
and the two Schur complements
\begin{align*}
\tM_{11}(\lambda) \eqbyd (M_{22}-\lambda I) - M_{12}'M_{11}^+ M_{12} \qquad 
\tM_{22}(\lambda) \eqbyd M_{11}-M_{12}(M_{22}-\lambda I)^+ M_{12}'
\end{align*}
The extended-value definition of $L(\cdot)$ in \cref{eq:Llam} is required in what follows. As established in \cref{prop:sdparalt}, $L(\lambda)$ is the optimal value of the parametric problems $\min_u\max_w L(u,w,\lambda)$ and $\max_w \min_u L(u,w,\lambda)$ only for $d \in R(M(\lambda))$; if $d \notin R(M(\lambda))$, neither problem has a solution and the optimal value is $+\infty$. The formula $-\half d'M^+(\lambda)d + \lambda/2$, by contrast, returns a \textit{finite} value for these $\lambda$ because the pseudoinverse discards the offending direction, and, as we shall see, that value is \textit{smaller} than the optimal value at neighboring $\lambda$. Omitting the second case of \cref{eq:Llam} therefore does not produce a visible failure in the optimizations over $\lambda$ appearing in \cref{prop:conminmaxalt}; it produces a wrong answer.

We next collect the properties of $L(\cdot)$ required to characterize those optimizations. Recall that $M(0) \geq 0$ implies $\tM_{11} \eqbyd \tM_{11}(0) \geq 0$, so that $\normf{\tM_{11}}$ is the largest eigenvalue of $\tM_{11}$, and that $\tM_{11}(\lambda) = \tM_{11} - \lambda I$.

\begin{lemma}[Properties of the dual function]
\label{lem:Ldual}
Given $M_{11}>0$ and $M(0) \geq 0$, let $L(\cdot)$ be the function defined in \cref{eq:Llam} and denote
\begin{equation*}
\td \eqbyd d_2 - M_{12}'M_{11}^{-1}d_1
\end{equation*}
\begin{enumerate}
\item \emph{(Reduction.)} For all $\lambda$, $d \in R(M(\lambda))$ if and only if $\td \in R(\tM_{11}(\lambda))$, and for such $\lambda$
\begin{equation}
L(\lambda) = - \half d_1'M_{11}^{-1}d_1 - \half \td'\tM_{11}^+(\lambda)\td + \lambda/2
\label{eq:Lreduced}
\end{equation}

\item \emph{(Derivatives and convexity.)} For $\lambda > \normf{\tM_{11}}$, matrix $M(\lambda)$ is nonsingular, the stationary point is unique and given by
\begin{equation*}
\begin{bmatrix}u^*(\lambda) \\ w^*(\lambda)\end{bmatrix} = -M^{-1}(\lambda)d \qquad w^*(\lambda) = -\tM_{11}^{-1}(\lambda)\td
\end{equation*}
and $L(\cdot)$ is infinitely differentiable with
\begin{align}
\frac{dL}{d\lambda}(\lambda) &= \half \left(1 - (w^*)'w^*(\lambda)\right) \label{eq:dLdlam}\\
\frac{d^2L}{d\lambda^2}(\lambda) &= \td'(\lambda I - \tM_{11})^{-3}\td = -(w^*)'\tM_{11}^{-1}(\lambda)w^*(\lambda) \geq 0 \label{eq:d2Ldlam}
\end{align}
so $L(\cdot)$ is convex on $(\normf{\tM_{11}}, \infty)$, and strictly convex wherever $w^*(\lambda) \neq 0$.

\item \emph{(Left endpoint.)} If $\td \notin R(\tM_{11}(\normf{\tM_{11}}))$, then $L(\normf{\tM_{11}}) = +\infty$ and $L(\lambda) \rightarrow \infty$ as $\lambda \searrow \normf{\tM_{11}}$. If $\td \in R(\tM_{11}(\normf{\tM_{11}}))$, then $L(\cdot)$ is finite and infinitely differentiable in a neighborhood of $\normf{\tM_{11}}$, and is therefore continuous and convex on $[\normf{\tM_{11}}, \infty)$.

\item \emph{(Coercivity.)} $L(\lambda) \rightarrow \infty$ as $\lambda \rightarrow \infty$.
\end{enumerate}
\end{lemma}

\begin{proof}\mbox{ }
\begin{enumerate}
\item Since $M_{11}>0$, we may multiply by a full rank matrix to obtain
\begin{equation*}
\begin{bmatrix} I & 0 \\ -M_{12}'M_{11}^{-1} & I \end{bmatrix}
\left( M(\lambda)x - d \right) =
\begin{bmatrix} M_{11} & M_{12} \\ 0 & \tM_{11}(\lambda) \end{bmatrix} x -
\begin{bmatrix} d_1 \\ \td \end{bmatrix}
\end{equation*}
so $M(\lambda)x=d$ has a solution if and only if $\tM_{11}(\lambda)x_2 = \td$ has a solution, and the first statement follows from \cref{prop:linalg}. Next consider the quadratic function $q(x) \eqbyd \half x'M(\lambda)x + x'd$ with $x=(u,w)$. If $d \in R(M(\lambda))$, every stationary point $x^*$ satisfies $M(\lambda)x^* = -d$ and therefore
\begin{equation*}
q(x^*) = -\half (x^*)'d = -\half d'M^+(\lambda)d
\end{equation*}
in which the last equality uses $x^* \in -M^+(\lambda)d + N(M(\lambda))$ and the fact that $d \in R(M(\lambda))$ is orthogonal to $N(M(\lambda))$ for symmetric $M(\lambda)$. Since $M_{11}>0$, we may partially minimize $q$ with respect to $u$, giving $u = -M_{11}^{-1}(M_{12}w + d_1)$ and
\begin{equation*}
\tilde q(w) = \half w' \tM_{11}(\lambda) w + w'\td - \half d_1'M_{11}^{-1}d_1
\end{equation*}
The stationary points of $q$ are exactly the pairs $(u(w), w)$ with $w$ stationary for $\tilde q$, and applying the argument above to $\tilde q$ gives $\tilde q(w^*) = -\half \td'\tM_{11}^+(\lambda)\td - \half d_1'M_{11}^{-1}d_1$, which establishes \cref{eq:Lreduced}.

\item Since $\tM_{11} \geq 0$, we have $\tM_{11}(\lambda) = \tM_{11}-\lambda I < 0$ for $\lambda > \normf{\tM_{11}}$, and $M(\lambda)$ is nonsingular by the factorization in part 1. The stationary point is therefore unique, and $w^* = -\tM_{11}^{-1}(\lambda)\td$ from part 1. Differentiating \cref{eq:Lreduced} with $S(\lambda) \eqbyd \tM_{11}(\lambda)$ and $dS/d\lambda = -I$ gives $dS^{-1}/d\lambda = S^{-2}$ and
\begin{equation*}
\frac{dL}{d\lambda} = -\half \td'S^{-2}(\lambda)\td + \half = \half\left(1 - (w^*)'w^*(\lambda)\right)
\end{equation*}
Differentiating a second time gives
\begin{equation*}
\frac{d^2L}{d\lambda^2} = -\td'S^{-3}(\lambda)\td = \td'(\lambda I - \tM_{11})^{-3}\td = -(w^*)'S^{-1}(\lambda)w^*(\lambda)
\end{equation*}
Since $S(\lambda) < 0$, we have $S^{-1}(\lambda) <0$ and $d^2L/d\lambda^2 \geq 0$ with equality if and only if $w^*(\lambda)=0$, which establishes convexity.

\item Let $\tM_{11} = \sum_i \mu_i q_iq_i'$ be a spectral decomposition, so that $\mu_i \leq \normf{\tM_{11}}$ for all $i$ and $\mu_i = \normf{\tM_{11}}$ for at least one $i$. From \cref{eq:Lreduced}, for $\lambda > \normf{\tM_{11}}$
\begin{equation}
L(\lambda) = -\half d_1'M_{11}^{-1}d_1 - \half \sum_i \frac{(q_i'\td)^2}{\mu_i - \lambda} + \lambda/2
\label{eq:Lspectral}
\end{equation}
The condition $\td \in R(\tM_{11}(\normf{\tM_{11}}))$ is precisely $q_i'\td = 0$ for every $i$ with $\mu_i = \normf{\tM_{11}}$. If it fails, the corresponding term of \cref{eq:Lspectral} tends to $+\infty$ as $\lambda \searrow \normf{\tM_{11}}$ while the remaining terms are bounded, so $L(\lambda) \rightarrow \infty$. If it holds, those terms are absent from \cref{eq:Lspectral}, the remaining terms have $\mu_i < \normf{\tM_{11}}$, and $L(\cdot)$ is a rational function with no pole at $\normf{\tM_{11}}$.

\item Every term of the sum in \cref{eq:Lspectral} tends to zero as $\lambda \rightarrow \infty$, and $\lambda/2 \rightarrow \infty$. \qedhere
\end{enumerate}
\end{proof}

Note that \cref{eq:dLdlam} shows that the outer optimization over $\lambda$ does nothing but enforce the constraint that was dualized: a stationary point of $L(\cdot)$ in the interior satisfies $(w^*)'w^* = 1$, i.e., $w^*(\lambda) \in W$. Note also that setting $M_{11}$, $M_{12}$, and $d_1$ to zero in \cref{eq:d2Ldlam} recovers the second derivative appearing in the proof of \cref{lem:dLquad}, with $(\tM_{11}, \td)$ in place of $(D, d)$.

\begin{subtheorem}{thm}
\setcounter{theorem}{20}
\setcounter{parentnumber}{\value{theorem}-1}
\end{subtheorem} 
\begin{subtheorem}{thm}
\begin{thm}[Alternative to 19 and 20]
\label{prop:conminmaxalt}
Consider quadratic function $V(\cdot): \bbR^{m+n} \rightarrow \bbR$, compact constraint set $W$, and 
Lagrangian function $L(\cdot): \bbR^{m+n+1} \rightarrow \bbR$
\begin{align*}
V(u,w) &= (1/2) \begin{bmatrix} u \\ w \end{bmatrix}'
M(0)
\begin{bmatrix} u \\ w \end{bmatrix}
+  \begin{bmatrix} u \\ w \end{bmatrix}' d
\qquad W \eqbyd \{ w \mid w'w = 1\} \\
L(u,w,\lambda) &= (1/2) \begin{bmatrix} u \\ w \end{bmatrix}'
{M(\lambda)}
\begin{bmatrix} u \\ w \end{bmatrix}  +
\begin{bmatrix} u \\ w \end{bmatrix}'  d
+  \lambda/2
\end{align*} 
We consider the two constrained optimization problems
\begin{alignat}{2}
&\min_u &\max_{w \in W} &V(u,w)  \quad \text{robust control} \label{eq:conminmaxalt}\\
&\max_{w \in W} &\min_u &V(u,w) \quad \text{worst-case feedforward control} \label{eq:conmaxminalt}
\end{alignat}
Assume $M(0) \geq 0$ and $M_{11} > 0$.
% Denote the Lagrangian function by
% \begin{equation*}
% L(\lambda) = -(1/2) d'M^+(\lambda)d + \lambda/2
% \end{equation*} 
For $d \in R(M(\lambda))$ denote stationary points by $(u^*(\lambda), w^*(\lambda))$ and recall the extended-value function $L(\cdot)$ defined in \cref{eq:Llam}
\begin{align*}
\begin{bmatrix}u^*(\lambda) \\ w^*(\lambda) \end{bmatrix} &\in -M^+(\lambda) d + N(M(\lambda)) \\
L(\lambda) &= V(u^*,w^*) -(1/2) \lambda((w^*)'w^* -1) = -(1/2) d'M^+(\lambda)d + \lambda/2 
\end{align*} 
with $L(\lambda) = +\infty$ for $d \notin R(M(\lambda))$.
We  then have the following results. 
\begin{enumerate}
\item
The solution to problem \eqref{eq:conminmaxalt} exists for all $d \in \bbR^{m+n}$ and is given by
\begin{equation*}
u_r^0 = u^*(\lambda_r^0) \qquad w_r^0 = \ow^0(u_r^0) \cap W
\end{equation*} 
where $\lambda^0_r$ denotes the solution to the following optimization, which exists for all $d \in \bbR^{m+n}$
\begin{equation}
 \lambda_r^0 = \arg \min_{\lambda \geq \norm{M_{22}}} L(\lambda)
\label{eq:lamr}
\end{equation} 
and $\ow^0(u_r^0)$, the solution of the inner maximization as in \cref{prop:sdparalt}, is all solutions to 
\begin{equation*}
M_{12}' u_r^0 + (M_{22} - \lambda_r^0 I) \ow^0(u_r^0) = -d_2
\end{equation*} 
The optimal cost is given by $V(u_r^0, w_r^0) = L(\lambda_r^0)$.

\item
The solution to problem \eqref{eq:conmaxminalt} exists for all $d \in \bbR^{m+n}$ and is given by
\begin{equation*}
u_f^0 = \uu^0(w_f^0) \qquad w_f^0 = w^*(\lambda^0_f) \cap W
\end{equation*} 
where $\lambda^0_f$ denotes the solution to the following optimization, which exists for all $d \in \bbR^{m+n}$
\begin{equation}
 \lambda_f^0 = \arg \min_{\lambda \geq \normf{\tM_{11}}} L(\lambda)
\label{eq:lamf}
\end{equation} 
and $\uu^0(w_f^0)$, the solution of the inner minimization as in \cref{prop:sdparalt}, is the unique solution to 
\begin{equation*}
M_{11} \uu^0(w_f^0) + M_{12} w_f^0 = -d_1
\end{equation*} 
The optimal cost is given by $V(u_f^0, w_f^0) = L(\lambda_f^0)$.

\end{enumerate}

\end{thm} 
\end{subtheorem}  

\begin{proof}
% First we show that $d \in R(M(\lambda))$ for $\lambda > \normf{\tM_{11}}$ so that we can invoke \cref{prop:sdparalt}.
% To this end, we seek $y$ such that $M(\lambda)y = d$ where $M(\lambda)$ is defined in \cref{eq:Mlam}.  The top half is
%  $M_{11} y_1 + M_{12} y_2 = d_1$, which has the unique solution $y_1 = M_{11}^{-1}(d_1 - M_{12} y_2)$ since $M_{11}>0$ by assumption.  Substituting $y_1$ into the bottom half, we then require $(M_{22}-\lambda I - M_{12}'M_{11}^{-1}M_{12})y_2 = d_2 - M_{12}'M_{11}^{-1} d_1$.  For $\lambda > \normf{\tM_{11}}$, this equation also has a unique solution $y_2$ for all $d$, and we have that $d  \in R(M(\lambda))$.  Note that since $\norm{M_{22}} \geq \normf{\tM_{11}} $, the range condition holds for $\lambda > \norm{M_{22}}$ as well.  Note that we do \textit{not} know that $d \in  R(M(\lambda))$ for $\lambda = \norm{M_{22}}$ if $\normf{\tM_{11}} = \norm{M_{22}}$ as depicted in \cref{fig:sdparalt}, top, case 2(b).

First note that $M(0)\geq 0$ implies that $M_{22}\geq 0$ and $\tM_{11} \geq 0$ as discussed after \cref{prop:psd}. 
\begin{enumerate}
\item First we note that from \cref{prop:conlag} problem \cref{eq:conminmaxalt} is equivalent to
\begin{equation*} 
 \min_u \max_w \min_\lambda L(u,w,\lambda) 
\end{equation*} 
Because $M_{22} \geq 0$, \cref{prop:conquad} applies, and we have strong duality in the inner two optimizations giving the equivalent problems
\begin{equation*} 
 \min_u \min_\lambda \max_w L(u,w,\lambda)  = \min_\lambda  \min_u \max_w L(u,w,\lambda)
\end{equation*} 
Now we apply \cref{prop:sdparalt} and insert the solution to the $\min_u \max_w L$ problem giving
\begin{equation*}
 \min_{\lambda \geq \norm{M_{22}}} L(\lambda) 
\end{equation*} 
which is \eqref{eq:lamr}, and $L(\cdot)$ is the extended-value function \cref{eq:Llam}. Note that the constraint on $\lambda$ follows because as stated in \cref{prop:sdparalt}, a solution to $\min_u\max_w L(u,w,\lambda)$ does not exist for $\lambda < \norm{M_{22}}$. The value $L(\lambda)=+\infty$ is likewise assigned exactly to those $\lambda \geq \norm{M_{22}}$ for which $\min_u\max_w L(u,w,\lambda)$ has no solution, which is case 2(b) of \cref{prop:sdparalt}.

Next we establish that a solution to \eqref{eq:lamr} exists for all $d \in \bbR^{m+n}$, and characterize it. Recall from the discussion after \cref{prop:sdparalt} that $\normf{\tM_{11}} \leq \norm{M_{22}}$, and consider the two mutually exclusive cases.
\begin{enumerate}
\item[(i)] $d \in R(M(\norm{M_{22}}))$.  By parts 2 and 3 of \cref{lem:Ldual}, $L(\cdot)$ is finite, continuous, and convex on $[\norm{M_{22}}, \infty)$, and coercive by part 4, so a solution to \eqref{eq:lamr} exists. Moreover, since $L(\cdot)$ is convex on this interval, the solution is the left end point, $\lambda_r^0 = \norm{M_{22}}$, if and only if $(d/d\lambda)L(\lambda) \geq 0$ at $\lambda = \norm{M_{22}}$, which by \cref{eq:dLdlam} is equivalent to $(w^*)'w^*(\norm{M_{22}}) \leq 1$. Otherwise $\lambda_r^0$ lies in the interior of the interval and satisfies $(w^*)'w^*(\lambda_r^0)=1$.

\item[(ii)] $d \notin R(M(\norm{M_{22}}))$. Here $M(\norm{M_{22}})$ is rank deficient, which as shown in the discussion following \cref{prop:sdparalt} is possible only if $\norm{M_{22}} = \normf{\tM_{11}}$. Part 3 of \cref{lem:Ldual} therefore applies at $\lambda=\norm{M_{22}}$ and gives $L(\lambda) \rightarrow \infty$ as $\lambda \searrow \norm{M_{22}}$. So $L(\cdot)$ is continuous on the open interval $(\norm{M_{22}}, \infty)$ and grows without bound at both ends, and a solution to \eqref{eq:lamr} exists and satisfies $\lambda_r^0 > \norm{M_{22}}$.
\end{enumerate}
Since every $d \in \bbR^{m+n}$ falls in case (i) or case (ii), a solution to \eqref{eq:lamr} exists for all $d$. Note that case (ii) is the exceptional one for \eqref{eq:lamr}: it requires $\norm{M_{22}} = \normf{\tM_{11}}$ \textit{and} $d \notin R(M(\norm{M_{22}}))$.

Given the optimal solution $\lambda^0_r$, we then evaluate the optimal $u^0_r$ and $w^0_r$ from the formulas in \cref{prop:sdparalt}, and we have to intersect the inner solution set $\ow^0(u_r^0)$ with set $W$ as in \cref{lem:SPlag} to obtain the solution to the original constrained problem \eqref{eq:conminmaxalt}.

\item Proceeding similarly, from \cref{prop:conlag} we note that problem \cref{eq:conmaxminalt} is equivalent to
\begin{equation*} 
\max_w \min_\lambda \min_u L(u,w,\lambda) 
\end{equation*} 
But here we must solve the inner $\min_u L$ if we wish to appeal to \cref{prop:conquad} for strong duality of the outer two problems. Since $M_{11}>0$, the inner problem has a unique solution for all $d \in \bbR^{m+n}$ with optimal value
 \begin{equation*}
L(u^0, w, \lambda) = (1/2) w'\tM_{11}w + w'(d_2 - M_{12}'M_{11}^{-1} d_1) - (1/2) d_1'M_{11}^{-1}d_1 -
\half \lambda (w'w - 1)
\end{equation*} 
Because $\tM_{11} \geq 0$, \cref{prop:conquad} again applies, and we have strong duality in the outer two problems giving the equivalent problem
\begin{equation*} 
\min_\lambda \max_w \min_u L(u,w,\lambda) 
\end{equation*} 
From here, we again apply \cref{prop:sdparalt} and proceed as in the previous part. We obtain the optimization problem
\begin{equation*}
 \min_{\lambda \geq \normf{\tM_{11}}} L(\lambda) 
\end{equation*} 
which is \eqref{eq:lamf}, and  the constraint on $\lambda$ follows because the solution to $\max_w\min_u L(u,w,\lambda)$ does not exist for $\lambda < \normf{\tM_{11}}$.

Existence of a solution to \eqref{eq:lamf} follows from the two cases of the previous part after replacing $\norm{M_{22}}$ by $\normf{\tM_{11}}$ throughout. The two cases do \textit{not}, however, retain their relative importance, and the argument of case (ii) above does not transfer. Since $\tM_{11} \geq 0$, the scalar $\normf{\tM_{11}}$ is the largest eigenvalue of $\tM_{11}$, so $\tM_{11}(\normf{\tM_{11}}) = \tM_{11} - \normf{\tM_{11}}I$ is singular and $M(\normf{\tM_{11}})$ is rank deficient for \textit{every} $M(0)$ satisfying the assumptions.  Case (ii), $d \notin R(M(\normf{\tM_{11}}))$, is therefore the generic case for \eqref{eq:lamf} and case (i) the exception, which is the reverse of the situation for \eqref{eq:lamr}. In case (ii) we again have $L(\lambda) \rightarrow \infty$ as $\lambda \searrow \normf{\tM_{11}}$ by part 3 of \cref{lem:Ldual}, a solution exists, and $\lambda_f^0 > \normf{\tM_{11}}$.  This behavior is worth emphasizing because the left end point is the value at which the pseudoinverse formula $-\half d'M^+(\lambda)d + \lambda/2$ is finite and \textit{smaller} than $L(\lambda_f^0)$; omitting the second case of \cref{eq:Llam} thus returns $\lambda = \normf{\tM_{11}}$ and a cost strictly below the true optimum.

Finally, in this problem, we must intersect the outer solution $w^*$ of $\max_w \min_u L$ with $W$, and evaluate the inner minimization $\uu^0(w_f^0)$ at the resulting $w_f^0$, to solve the original constrained problem \eqref{eq:conmaxminalt}. \qedhere
\end{enumerate}  
\end{proof} 
 
Note that strong duality does \textit{not} hold in general for problems \cref{eq:conminmaxalt} and \cref{eq:conmaxminalt}: worst-case feedforward control achieves lower (not higher) cost than robust control, which makes sense since the controller has knowledge of the disturbance $w$ in the feedforward problem, but not the robust control problem.  As shown in \cref{fig:sdparalt}, the final $\min_{\lambda}L(\lambda)$ takes place over a strictly larger set in the feedforward problem when $\normf{\tM_{11}} < \normf{M_{22}}$, allowing strictly lower cost for worst-case feedforward control compared to robust control.

\paragraph{The inequality constrained problem.}

Returning to the original constrained problem for a moment, adopt the definitions of $V$, $W$, $\lambda$, and $L$ in \cref{prop:conquad}
\begin{gather*}
V(w) \eqbyd \half w'Dw + w'd \qquad 
L(w, \lambda) \eqbyd V(w) - \half \lambda (w'w-1)
\end{gather*} 
with $D \geq 0$, and consider the following modified problem in which the equality constraint is replaced by an inequality
\begin{equation}
 \max_{w \in W_f} V(w) \qquad W_f \eqbyd \{w \mid w'w \leq 1\}
\label{eq:maxWf}
\end{equation} 
The Lagrangian reformulation is
\begin{equation}
 \max_w \inf_{\lambda \geq 0} L(w, \lambda)
\label{eq:maxminLf}
\end{equation} 
Denote the inner problem of \eqref{eq:maxminLf} by $\oL(w) \eqbyd \inf_{\lambda \geq 0} L(w,\lambda)$ and note that 
\begin{equation*}
\oL(w) = \begin{cases} V(w), \quad &w \in W_f \\
 -\infty, \quad &w \notin W_f 
\end{cases} 
\end{equation*}   
The three cases behave differently. For $w'w<1$ the infimum is attained at the unique minimizer $\lambda^0(w) = 0$. For $w'w=1$ the function $L(w,\cdot)$ is constant, so the infimum is attained and $\lambda^0(w) = [0,\infty)$.  For $w \notin W_f$ we have $w'w-1>0$, so $L(w,\lambda) \rightarrow -\infty$ as $\lambda \rightarrow \infty$ and the infimum is $-\infty$ and is \textit{not} attained, which is why \eqref{eq:maxminLf} is written with an infimum.  Since the solution of \eqref{eq:maxWf} exists, $\max_{w \in W_f} V(w) = \max_w \oL(w)$, so the Lagrangian reformulation \eqref{eq:maxminLf} has the same solution as \eqref{eq:maxWf}.

Next consider the dual Lagrangian problem
\begin{equation}
\min_{\lambda \geq 0}  \max_w L(w, \lambda)
\label{eq:minmaxLf}
\end{equation} 
and we establish strong duality of problems \eqref{eq:maxminLf} and \eqref{eq:minmaxLf}.

\begin{proposition}[Strong duality of inequality constrained optimization]
\label{prop:ineqdual}
Solutions to \eqref{eq:maxWf}, \eqref{eq:maxminLf}, and \eqref{eq:minmaxLf} exist for all $D \geq 0$ and $d \in \bbR^n$, and 
problems \eqref{eq:maxminLf} and \eqref{eq:minmaxLf} satisfy strong duality.
\end{proposition} 
\begin{proof}
The solution to \eqref{eq:maxWf} exists since $W_f$ is compact and $V(\cdot)$ is continuous. The argument above establishes that \eqref{eq:maxminLf} has the same solution, and we next prove that strong duality holds and that \eqref{eq:minmaxLf} also has the same solution. 
Note that weak duality always holds so that
\begin{equation}
 \max_w \inf_{\lambda \geq 0} L(w, \lambda) \leq \min_{\lambda \geq 0} \max_w L(w, \lambda)
\label{eq:weakLf}
\end{equation} 
So we establish the reverse inequality, which gives equality and hence strong duality. From \cref{lem:dLquad}, item 1, we have that
\begin{equation*}
 \min_\lambda \max_w L(w, \lambda) = \min_{\lambda \geq \norm{D}} \max_w L(w, \lambda) \geq \min_{\lambda \geq 0} \max_w L(w,\lambda)
\end{equation*} 
in which the last inequality follows because $\norm{D} \geq 0$, so the minimization on the right is over a larger set.  But since we also know that
\begin{equation*}
 \min_\lambda \max_w L(w, \lambda) \leq \min_{\lambda \geq 0} \max_w L(w,\lambda)
\end{equation*} 
for the same reason, we have that 
\begin{equation*}
 \min_\lambda \max_w L(w, \lambda) = \min_{\lambda \geq 0} \max_w L(w, \lambda)
\end{equation*} 
and, in particular, the solution to \eqref{eq:minmaxLf} exists because the solution to the unconstrained $\lambda$ problem exists by \cref{lem:dLquad}. Using the strong duality of the unconstrained $\lambda$ problem from \cref{prop:conquad} in this equality gives 
\begin{equation*}
\max_w  \min_\lambda L(w, \lambda) = \min_{\lambda \geq 0} \max_w L(w, \lambda)
\end{equation*} 
Again, because the inner minimization is over a smaller set, we also have
\begin{equation*}
\max_w \inf_{\lambda \geq 0} L(w, \lambda) \geq \max_w \min_\lambda L(w, \lambda)
\end{equation*} 
and using this result in the previous equation gives
\begin{equation*}
\max_w \inf_{\lambda \geq 0} L(w, \lambda) \geq  \min_{\lambda \geq 0} \max_w L(w, \lambda) 
\end{equation*} 
which is the desired reverse inequality of \eqref{eq:weakLf}, strong duality is established, and the proof is complete.
\end{proof} 

\appendix
\section*{Changes between versions}

Theorem numbering is stable across the three versions: a result added in a later version takes a number that an earlier version left vacant, so that a citation of a numbered result resolves to the same statement in every version.

The \textbf{second version} provided the following corrections and extensions of the first version.
 \begin{enumerate}
 \item The optimal $u$ and $w$ formulas in the original Corollary 13, Proposition 14, Corollary 19, and Proposition 20 were corrected.

\item Corollary 13 and Proposition 14 were combined in \cref{prop:sdparalt}. 
\item Corollary 19 and Proposition 20 were combined in \cref{prop:conminmaxalt}.

\item \cref{prop:SPalt} is a generalization of the previous Proposition 12.

\item Propositions \ref{prop:mina}, \ref{prop:minb}, and \ref{prop:minc} were new in that version.

 \item \cref{fig:sdparalt} was revised to illustrate \cref{prop:sdparalt}.
\end{enumerate} 

This \textbf{third version} corrects errors in the second version and adds the new results listed below.  The numbering of the second version is unchanged, and the new results occupy numbers that the second version left vacant.  Readers who have used the second version should note the first two items in particular.
 \begin{enumerate}
\item \textit{The two formulas for the inner optimizations in \cref{prop:conminmaxalt} were erroneously interchanged.}  Both are now written as in \cref{prop:sdparalt}.

\item \textit{The function $L(\lambda)$ is now defined to be $+\infty$ where $d \notin R(M(\lambda))$}, and the optimizations \eqref{eq:lamr} and \eqref{eq:lamf} are to be read with that convention.  At such $\lambda$ the underlying minmax problem has no solution, but the formula $-\half d'M^+(\lambda)d + \lambda/2$ returns a finite value that is smaller than the true optimum, so the previous statement of \cref{prop:conminmaxalt} returns a spurious answer rather than an evident failure.  This affects \eqref{eq:lamf} generically, since $M(\normf{\tM_{11}})$ is rank deficient for every $M(0)$.

\item \cref{lem:Ldual} is new. It establishes the derivatives and the convexity of $L(\cdot)$, which the boundary test in \cref{prop:conminmaxalt} requires and which does not follow from the corresponding argument in \cref{lem:dLquad}.

\item \cref{prop:ineqdual} is new. It treats the inequality constrained problem $\max_{w'w \leq 1} V(w)$, and establishes strong duality of its Lagrangian and dual Lagrangian formulations.

\item The proof of \cref{prop:conminmaxalt} now establishes that the optimizations \eqref{eq:lamr} and \eqref{eq:lamf} have solutions for all $d$, which the second version asserted without proof, and gives the test that decides whether the solution lies at the left end point of the interval or in its interior.

\item The discussion following \cref{prop:sdparalt} is new.  It establishes that case 2(b) is incompatible with cases 4(a) and 4(b): if $\normf{\tM_{11}} < \norm{M_{22}}$, then $M(\norm{M_{22}})$ is full rank and no $d$ satisfies case 2(b).

\item The Remark following \cref{prop:minc} has been removed.  Its claim that $M^+=X$ is false (in general), and the pseudoinverse question it raised is a distraction from the purpose of this note.

\item The example following \cref{prop:sdparalt} has been replaced, because the previous one satisfied neither $M_{11}>0$ nor $M(0) \geq 0$.

\item Typo corrections local to a proof, leaving the statements unchanged: the sign of $w^*$ and the existence of a unit-norm $w^*$ in \cref{lem:SPlag}; the necessity of $d \in R(M)$ and the range condition in \cref{prop:SPalt}; the case split, a determinant, and a stray factor in \cref{lem:dLquad}; and the citation of the saddle-point theorem in \cref{lem:Lquad}.

\item \cref{prop:conlag} is stated for a general continuous $\rho(\cdot)$ vanishing exactly on the constraint set, rather than for a distance function, and the maximization counterpart is added, because that is the form used in \cref{lem:Lquad} and in both parts of \cref{prop:conminmaxalt}.

\item A brief summary of the literature leading to the minimax theorem, \cref{th:minimax}, has been added.  Further connections between \cref{prop:conquad} and the fields of numerical linear algebra and recent results on the trust-region problem have also been added, and the discussion of the novelty of \cref{prop:conquad} is modified accordingly. 
\end{enumerate}

\bibliographystyle{abbrvnat}
\bibliography{abbreviations,books,articles,resgrpreports,proceedings}

\end{document}